\documentclass[aps,pre,showpacs,floats,floats,twocolumn,superscriptaddress,floatfix,english]{revtex4-1}

\usepackage[T1]{fontenc}
\usepackage[latin9]{inputenc}
\usepackage{amsmath}
\usepackage{amssymb}
\usepackage{graphicx}

\usepackage{graphicx}
\usepackage{epstopdf, epsfig}
\usepackage{dutchcal}
\usepackage{array}
\usepackage{multirow}
\usepackage{setspace}
\usepackage{subfigure}
\usepackage{caption}
\usepackage{color}
\usepackage{epstopdf}
\usepackage[normalem]{ulem}
\usepackage{float}
\usepackage{chemformula}
\usepackage{siunitx}

\DeclareMathAlphabet\mathzapf{T1}{pzc}{mb}{it}
\DeclareMathAlphabet\mathrsfso{U}{rsfso}{m}{n}

\newcommand{\bb}{\boldsymbol}

\DeclareMathOperator{\Log}{Log}

\def\XXint#1#2#3{{\setbox0=\hbox{$#1{#2#3}{\int}$}
     \vcenter{\hbox{$#2#3$}}\kern-.5\wd0}}

\newcommand{\step}{\Theta} 
\newcommand{\cauchy}{\mathzapf{P}} 
\newcommand{\dair}{\rho_{\rm{a}}} 
\newcommand{\dwat}{\rho_{\rm{w}}} 
\newcommand{\zc}{z_{\rm{c}}} 
\newcommand{\crit}{\chi_{\rm{c}}} 
\newcommand{\Usim}{\mathfrak{U}} 
\newcommand{\xic}{\xi_{\rm{c} }} 

\newcommand{\Z}{\mathzapf{Z}} 

\newcommand{\Us}{U_{\rm{S}}} 
\newcommand{\Lair}{L_{\rm{a}}} 
\newcommand{\Lwat}{L_{\rm{w}}} 
\newcommand{\Vair}{V_{\rm{a}}} 
\newcommand{\Vwat}{V_{\rm{w}}} 

\newcommand{\da}{d_{\rm{a}}} 
\newcommand{\dw}{d_{\rm{w}}}  
\newcommand{\uair}{u_{\star\rm{a}}} 
\newcommand{\uwat}{u_{\star\rm{w}}}  
\newcommand{\za}{z_{0\rm{a}}} 
\newcommand{\zw}{z_{0\rm{w}}}  

\newcommand{\g}{\mathcal{G}} 
\newcommand{\s}{\mathcal{S}} 

\newcommand{\z}{\mathcal{z}} 
\newcommand{\kd}{\mathcal{k}} 
\newcommand{\U}{\mathcal{U}} 
\newcommand{\Upc} {\U'_{\rm{c}}}  
\newcommand{\Uppc} {\U''_{\rm{c}}}  
\newcommand{\C}{\mathcal{C}} 
\newcommand{\short}{\epsilon} 

\newcommand{\st}{\tilde{\s}} 
\newcommand{\gt}{\tilde{\g}} 
\newcommand{\A}{\mathcal{A}} 
\newcommand{\B}{\mathcal{B}} 
\newcommand{\Rone}{\mathcal{R}_1} 
\newcommand{\Rtwo}{\mathcal{R}_2} 

\newcommand{\zcd}{\mathcal{z}_{\rm{c}}} 

\newcommand{\zcpm}{\mathcal{z}_{\rm{c}\pm}} 
\newcommand{\zcp}{\mathcal{z}_{\rm{c}+}} 
\newcommand{\zcm}{\mathcal{z}_{\rm{c}-}} 

\newcommand{\F}{F_{\rm{in}}} 
\newcommand{\f}{f_{\rm{out}}} 
\newcommand{\funif}{f_{\rm{unif},0}} 
\newcommand{\fl}{\f^{\ell}} 
\newcommand{\fu}{\f^{\rm{u}}} 
\newcommand{\fln}{f_1^{\ell}} 
\newcommand{\fun}{f_1^{\rm{u}}} 
\newcommand{\funiff}{f_{\rm{unif},1}} 
\newcommand{\zl}{\z_1^{\ell}} 
\newcommand{\zu}{\z_1^{\rm{u}}} 
\newcommand{\euler}{\gamma_{\rm{E}}} 
\newcommand{\zt}{\z_{\star}} %

\newcommand{\be}{\begin{equation}}
\newcommand{\ee}{\end{equation}}
\newcommand{\bc}{\begin{cases}}
\newcommand{\ec}{\end{cases}}

\newcommand{\nordita}{Nordita, KTH Royal Institute of Technology and Stockholm University, Stockholm 10691, Sweden}

\begin{document}

\title{Flow driven interfacial waves: an inviscid asymptotic study}

\author{A. F. Bonfils}
\affiliation{\nordita}
\author{Dhrubaditya Mitra}
\affiliation{\nordita}
\author{W. Moon}
\affiliation{ Department of Environmental Atmospheric Sciences, Pukyong National University, 48513 Pusan, South Korea }
\author{J. S. Wettlaufer}
\affiliation{\nordita}
\affiliation{Yale University, New Haven, Connecticut 06520-8109, USA}

\makeatother

\begin{abstract}
Motivated by wind blowing over water, we use asymptotic methods to study the evolution of short wavelength interfacial waves driven by the combined action of these flows. We solve the Rayleigh equation for the stability of the shear flow, and construct a uniformly valid approximation for the perturbed streamfunction, or eigenfunction. We then expand the real part of the eigenvalue, the phase speed, in a power series of the inverse wavenumber and show that the imaginary part is exponentially small. We give expressions for the growth rates of the \citet{miles57} and rippling \cite[e.g.,][]{young-wolfe} instabilities that are valid for an arbitrary shear flow. The accuracy of the results is demonstrated by a comparison with the exact solution of the eigenvalue problem in the case when both the wind and the current have an exponential profile. 
\end{abstract}
\maketitle

\section{Introduction}
\begin{figure}
  \centerline{\includegraphics[scale=0.2]{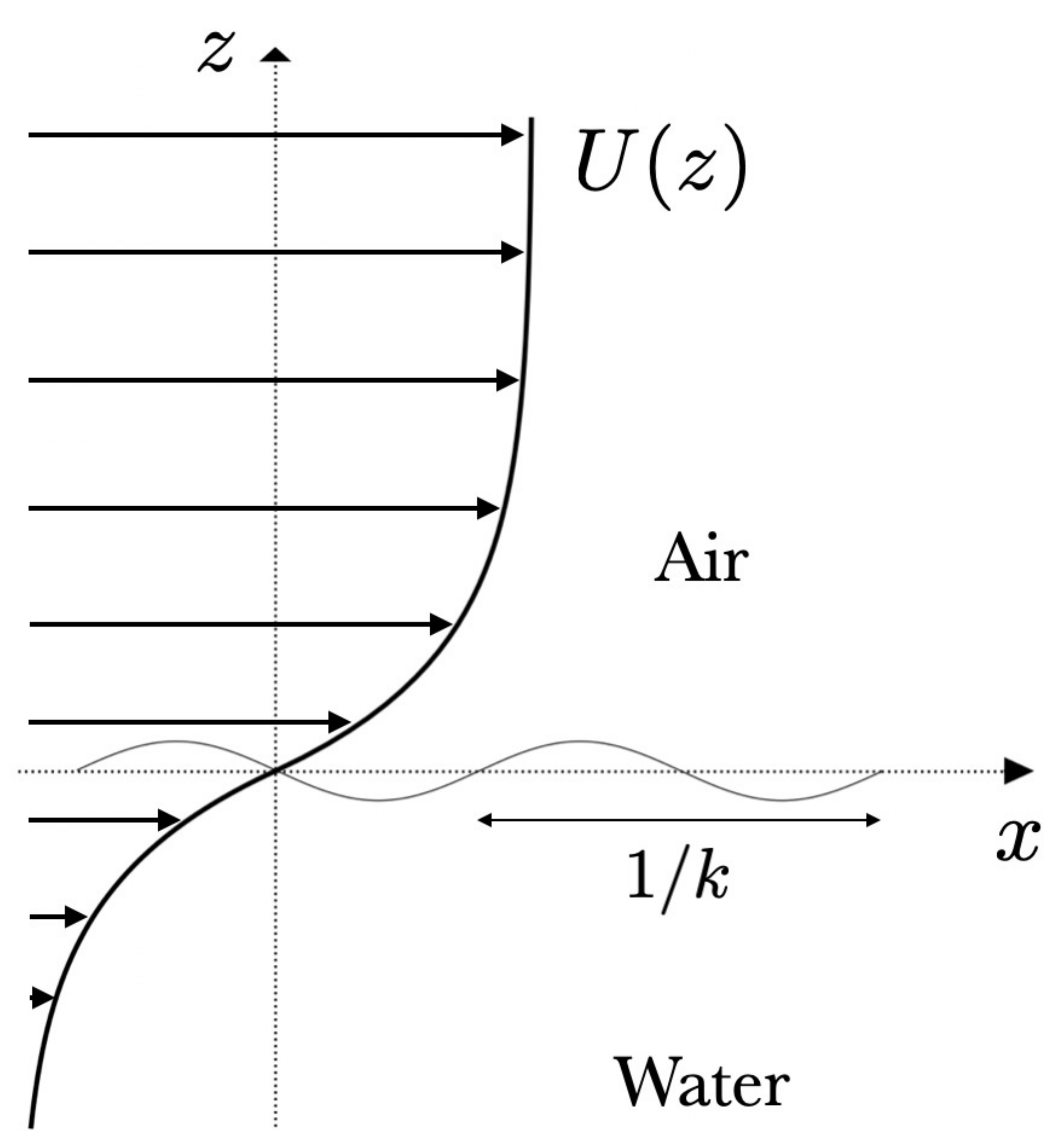}}
  \caption{Schematic of wind driving a current in the water. Both are modeled as a parallel shear flow $U=U(z)$ and perturbed by a wave with wavenumber $k$.}
\label{fig1}
\end{figure}
Waves at the interface between two fluids with different densities are ubiquitous in nature.  A natural question concerns how a flow in either fluid affects these waves? Here, we consider fluid layers of infinite extent. A canonical example is the wind flowing over the ocean, both of which can be modeled as parallel flows of the form $\bb{U}=U(z)\ \bb{\hat{x}}$, with $z$ the vertical coordinate and $\bb{\hat{x}}$ a horizontal unit vector (Figure \ref{fig1}). The stability of $\bb{U}$ under the influence of small two-dimensional perturbations has been studied extensively over the last seventy years. In the absence of a current in the water, \citet{miles57} found an instability of the wind leading to the growth of water waves. The theory of Miles is inviscid and assumes that the wind and the waves are weakly coupled because the air/water density ratio, $r$, is small. Hence, the wind transfers energy to the waves within a critical layer in the air, where the wind speed equals the phase speed of free surface waves. \textcolor{black}{Recent laboratory measurements of the air flow over wind-generated waves provide evidence of this critical layer mechanism \citep{carpenter-buckley-veron}. Nonetheless,} the growth rate can be calculated analytically in only a few cases \citep{bonfils-et-al22}.

 \textcolor{black}{\citet{miles57} argued that the critical layer should be above the viscous sublayer, and hence neglected viscosity. This rationale is now supported by the measurements of \citet{carpenter-buckley-veron}. The experiments of \citet{caulliez} showed that viscous damping is the main dissipation mechanism for waves shorter than 4 cm whereas at larger wavelengths the generation of capillary waves, micro-breaking and breaking also contribute to dissipation. Recent fully coupled direct numerical simulations further demonstrated that the viscous stress plays a role on wave growth only in the case of strong wind forcing~\citep{fullDNS2022}. Hence, the effect of viscosity is complex and has yet to be clarified \citep{zeisel-et-al,wu-deike}. Thus, simplified inviscid models still provide valuable insights, as shown here.  }

When there is a laminar current in the water, the position of the critical layer responsible for the Miles instability is unknown because the phase speed is itself unknown. This is associated with the fact that the surface waves are no longer free in the sense that the current modifies their dispersion relation in a non-trivial manner. Water currents are often wind-induced and decay with depth. \textcolor{black}{\citet{stern-adam} were the first to suggest that in such cases, sheared surface waves propagating slower than the water surface, which is dragged by the wind, undergo an instability. They considered a current with a broken-line velocity profile, a model further studied by \citet{caponi-et-al}, and extended to smooth velocity profiles by \citet{morland-et-al}. \citet{young-wolfe} showed that there is another critical layer in the water, where the unknown phase speed of the waves matches the speed of the current. They referred to this phenomenon as the "rippling instability".} Analytical progress on the rippling instability in deep water has only been made for piece-wise linear or exponential currents; see \citet{young-wolfe} for a review. Finally, \citet{anubhab} gave an exact analytical treatment of the stability of an exponential wind profile over a finite-depth water layer that is either quiescent or has linear or quadratic current profiles. For the quadratic current, they introduced spheroidal wave functions to assess the stability of a shear flow. Here, however, we obtain general results on both the Miles and rippling instabilities using asymptotic analysis. Here, we use asymptotic methods to obtain general results on both the Miles and the rippling instabilities. The fully coupled numerical approach of \citet{fullDNS2022} resolved the development of the shear-induced drift layer beneath the water surface as well as the evolution of the air-side turbulent boundary layer. They showed that in strongly forced cases the flows are transient, in the sense that the waves feedback on the air flow while the water flow becomes turbulent. Although transient effects are beyond the scope of this work, we can explicitly account for the effect of turbulence using mean profiles.

In the absence of a current in the water and for $r$ less than unity, and yet not small, the Miles instability is difficult to study. Indeed, the waves are then strongly coupled to the wind so that the shear substantially changes the dispersion relation, yielding yet another situation in which the position of the critical layer is unknown. Not only is such a strong wind--wave coupling a central process on Earth, but it may play an important role in a number of astrophysical settings, including white dwarfs, one of the end products of stellar evolution, neutron stars and black holes~\citep{shapiro2008black}.  For example, although most recent high-resolution three-dimensional simulations~\citep{casanova2011kelvin} suggest that Kelvin-Helmholtz instabilities driven by buoyant fingering may be able to explain the composition of white dwarfs, an alternative proposal by \citet{rosner2001} and \citet{alexakis2004heavy}, is that the Miles instability is responsible.  

Because most ocean waves have wavelengths much larger than the characteristic length scale of the wind profile, the Miles instability can be treated using asymptotic methods in the long-wave-limit~\citep{bonfils-et-al22}. Here, we focus on short waves with the goal of capturing the combined influences of an underlying current and a moderate density ratio. Whereas short waves may not be central in terrestrial oceanography, \citet{alexakis-et-al04} argued that they are at the core of the astrophysical setting described in the previous paragraph. \textcolor{black}{White dwarfs are extremely dense and dim objects, with many solar masses confined within Earth scale radii, and hence possess large gravitational forces. Indeed, \citet{alexakis-et-al04} showed that for an exponential wind profile, the low wavenumber cut-off of the Miles instability is a growing function of gravity, and hence the low wavenumber cut-off of the Miles instability is in fact very large. Thus, the growing waves at the surface of white dwarfs are short with respect to astrophysical scales.}

Therefore, rather than solving a particular geophysical or astrophysical problem, we treat a basic fluid mechanical question: the stability of a sheared two-fluid interface where the upper fluid is less dense than the lower fluid. For clarity of discussion, we refer to the
upper and lower fluids as air and water, respectively, and refer to wind as the shear flow in the air, and current as the shear flow in the water.

In \S \ref{stab}, we describe the linear stability analysis of a sheared two-fluid interface.  The eigenfunction satisfies the Rayleigh equation and the eigenvalue is a complex intrinsic phase speed.  We focus on the case of a wind-induced current in \S \ref{shortwave},  and obtain the real part of the eigenvalue as a series in powers of the dimensionless inverse wavenumber. Moreover, we show that the imaginary part of the eigenvalue is small and obtain general formulae for the growth rates of the Miles and the rippling instabilities. In \S \ref{discussion}, we treat the case of the wind and the current governed by an exponential profile and compare our asymptotic results to the exact eigenvalue.  In \S \ref{doublecrit}, we treat another type of current wherein a mode can have two critical layers, one in the air and one in the water. 
We then derive in \S \ref{asympsol} the asymptotic solution of the Rayleigh equation that was used in \S \ref{shortwave} for the calculation of the eigenvalue. In particular, we show that an internal boundary layer emerges from the singularity at the critical level. Finally, before concluding, in~\S \ref{comment} we demonstrate the limitations of the Wentzel-Kramers-Brillouin (WKB) approach to solving the Rayleigh equation.

\section{Linear stability of a sheared two-fluid interface}\label{stab}


\citet{young-wolfe} derived the eigenvalue problem associated with the linear stability of an inviscid parallel shear flow across a two-fluid interface as sketched in Fig.~\ref{fig1}.  First we outline the governing Rayleigh equation and boundary conditions, after which we describe the non-dimensionalization of the problem. 

\subsection{Eigenvalue problem} \label{eigenpb}

We consider a parallel shear flow, $U=U(z)$, monotonic in both air and water \textcolor{black}{with a non-zero curvature}. The air to water density ratio is $r\equiv \dair/\dwat<1$, and the gravitational acceleration and surface tension are $g$ and $\sigma$ respectively.
Incompressibility ensures that a perturbation of the flow is entirely determined by the streamfunction $\psi=\psi(x,z,t)$, where $t$ is the time. We use normal modes in the form
\be
\psi(x,z,t)=\Re\big\{ \hat{\psi}(z)\ e^{ik(x-ct)} \big\},
\ee
where $k$ is a real wavenumber and $c$ a complex phase speed to be determined (the eigenvalue), conservation of vorticity yields the Rayleigh equation as
\be
\big[U(z)-c\big]\big[\hat{\psi}''(z)-k^2\hat{\psi}(z)\big] - U''(z) \hat{\psi}(z)= 0. \label{Rayleigh}
\ee
We require the function $U(z)$ to be continuous at $z=0$, which excludes a Kelvin-Helmholtz type of instability and ensures the continuity of $\hat{\psi}(z)$ \citep{drazin-reid}.  Note that the derivative $U'(z)$ may have a finite jump at $z=0$ and the perturbation must decay in the far field. Provided that 
\be
\lim\limits_{z\to\pm\infty} \frac{U''(z)}{U(z)-c} =0,
\ee 
we can impose an exponential decay, viz.,
\be
\lim\limits_{z\to\pm\infty} \psi'(z)\pm k \psi(z) =0. \label{BCinf} 
\ee
Finally, we impose continuity of the normal stress at $z=0$ and require the air-water interface to be a streamline, which yields the boundary condition
\begin{widetext}
\be
\frac{\sigma}{\dwat}\ k^2 + (1-r) g + r \Big[ (\Us- c)^2 \frac{\hat{\psi}'(0^+)}{\hat{\psi}(0)} -  (\Us- c) U'(0^+)\Big] - (\Us- c)^2 \frac{\hat{\psi}'(0^-)}{\hat{\psi}(0)} + (\Us- c) U'(0^-)=0, \label{BCdim}
\ee
\end{widetext}
where $\Us= U(z=0)$ is the surface drift. 

 \subsection{Non-dimensionalization} 
 
 The length and velocity scales, $L$ and $V$, differ in the air and water, which we denote with the subscripts $\rm{a}$ and $\rm{w}$ respectively. Whereas when there is no current in the water, we use the scales of the wind, $\Lair$ and $\Vair$, to non-dimensionalize the problem, because water is the primary medium of surface waves we use $\Lwat$ and $\Vwat$ when a current is present. Thus, the ratios 
\refstepcounter{equation}
$$
\Rone \equiv \frac{ \Vair}{ \Vwat}\qquad\text{and}\qquad \Rtwo \equiv \frac{ \Lair}{ \Lwat} 
  \eqno{(\theequation{\mathit{a},\mathit{b}})}
$$
act as additional control parameters. In a frame moving at speed $\Us$, we use the dimensionless profile
\be
\U\equiv \frac{U-\Us}{\Vwat}\ ,\qquad\text{where}\qquad \U(0)=0,
\ee
and the dimensionless intrinsic phase speed is
\be
\C\equiv \frac{c-\Us}{\Vwat}\ . 
\ee
The dimensionless wavenumber and vertical coordinate are $\kd \equiv k\Lwat$ and $\z\equiv z/\Lwat$, respectively, and we define the dimensionless gravity and surface tension as
\be
\g \equiv \frac{g\Lwat}{\Vwat^2}\qquad\text{and}\qquad \s \equiv \frac{\sigma}{\dwat \Vwat^2 \Lwat}\ .
\ee
From equation (\ref{BCinf}), the far field behavior of the streamfunction has the form $e^{\pm \kd\z}$. Short waves are characterized by $\kd\gg1$, so the exponential decay is captured as follows, 
\be
\frac{\hat{\psi}(z)}{\hat{\psi}(0)} \equiv
\bc
e^{-\kd \z} f(\z)\qquad\text{if}\qquad \z>0,\\
e^{\kd \z}\ h(\z)\qquad\text{if}\qquad \z<0. 
\ec\label{Transf}
\ee
We introduce the small parameter $\short\equiv1/\kd$ and use (\ref{Transf}) in the Rayleigh equation (\ref{Rayleigh}), which gives
\begin{align}
\short f''(\z)- 2 f'(\z)- \short\ \frac{\U''(\z)} {  \U(\z)-\C }\ f(\z)&=0, \quad  f(0)=1, \label{feq}\\
\text{and}\qquad\short h''(\z)+2 h'(\z)- \short\ \frac{\U''(\z)} {  \U(\z)-\C }\ h(\z)&=0, \quad  h(0)=1.  \label{heq}
\end{align}
We assume that 
\refstepcounter{equation}
$$
f(\z)=O(1)\quad\text{as}\quad \z\to+\infty\quad\text{and}\qquad h(\z)=O(1)\quad\text{as}\quad \z\to-\infty, \label{newBCinf}
  \eqno{(\theequation{\mathit{a},\mathit{b}})}
$$
the veracity of which we check a posteriori. The dimensionless form of equation (\ref{BCdim}) is
\begin{widetext}
\be
\frac{\s}{ \short} + \g(1-r)\short +  \big[ r \U'(0^+)-\U'(0^-)\big] \short \C  - (1+r) \C^2 +   \big[ r f'(0^+)-h'(0^-)\big]\short\C^2 =0.\label{BC}
\ee
\end{widetext}
 For a given profile $\U(\z)$, our main task is to solve equations (\ref{feq}) and (\ref{heq}) subject to the boundary conditions (\ref{newBCinf}) and (\ref{BC}). 

We consider the canonical situation where the wind blows over the water in which it induces a current (fig. \ref{fig1}). Hence, $\U'\big(\z\lessgtr0\big)>0$, $\U(\z>0)>0$, and $\U(\z<0)<0$. We explore another situation in \S \ref{doublecrit}. 

\subsection{Examples of profiles} \label{profiles}

A typical example of a wind and a wind-induced current is the double exponential profile,
\be
U(z)= 
\bc
U_\infty + (\Us- U_\infty)\ e^{-z/\da} \qquad\text{if}\qquad z>0,\\
\Us\ e^{z/\dw}\qquad\text{if}\qquad z<0,
\ec\label{doubleexp}
\ee
where $U_\infty$ is the free-stream air velocity, and $\da$ and $\dw$ denote the thicknesses of the air and water shear boundary layers respectively.  In the frame of the water surface, the dimensionless form of Eq. (\ref{doubleexp}) is
\be
\U(\z)= 
\bc
 (\Rone- 1)(1-\ e^{-\z/\Rtwo}) \qquad\text{if}\qquad \z>0,\\
e^{\z}-1\qquad\text{if}\qquad \z<0,
\ec
\ee
with $\Rone= U_\infty/\Us$ and $\Rtwo= \da/\dw$. \citet{young-wolfe} showed that for such a profile the eigenvalue problem can be solved exactly in terms of hypergeometric functions. 

In the context of physical oceanography, the double log profile, 
\be
U(z)= 
\bc
\Us+\uair \ln(1+z/\za)/\kappa  \qquad\text{if}\qquad z>0,\\
\Us -\uwat \ln(1-z/\zw)/\kappa   \qquad\text{if}\qquad z<0,
\ec\label{doublelog}
\ee
may be more realistic \citep{wu75}, where $\uair$ and $\uwat$ are the friction velocities of air and water, respectively, $\kappa=0.4$ is the von K\'arm\'an constant, and $\za$ and $\zw$ are air and water roughness lengths, accounting for the presence of waves. Note that the velocity of the logarithmic current is negative for $z<z_{\rm{min}}$, where
\be
z_{\rm{min}}\equiv -\zw \big(e^{\kappa\Us/\uwat}-1\big).
\ee
Such a change of sign is unphysical so we take $U(z)=0$ for $z<z_{\rm{min}}$. The dimensionless form of (\ref{doublelog}) in the frame of the water surface is
\be
\U(\z)= 
\bc
\Rone\ln(1+\z/\Rtwo)/\kappa \qquad\text{if}\qquad \z>0,\\
-\ln(1-\z)/\kappa\qquad\text{if}\qquad \z<0,
\ec
\ee
with $\Rone= \uair/\uwat$ and $\Rtwo= \za/\zw$. In this case, exact analytical solutions are unknown. 

\section{Short wavelength expansions and exponential asymptotics} \label{shortwave}

We treat the eigenvalue problem described in \S\ref{eigenpb} perturbatively, where the small parameter is the dimensionless inverse wavenumber, $\short\ll1$.  We draw intuition for the approach from \citet{miles57}, who considered a simpler version of our problem, with the small parameter $r\ll1$ and in the absence of a current. Hence, the leading order eigenvalue corresponding to $r=0$ is real and equal to $c_s$, the phase speed of free surface waves. Moreover, due to the critical layer at height $z_c$, such that $U(z_c)=c_s$, the eigenvalue has an imaginary part of order $r$ as well as real corrections, also of order $r$. 

In contrast to the treatment of \citet{miles57}, the  leading order eigenvalue is unknown.   In fact, even the nature of the lowest order behavior in $\short$ is murky and hence we only assume that 
\be
\C = \C_r(\short) + \rm{i}\ \C_i(\short), \qquad\text{with}\qquad \C_i(\short)\ll \C_r(\short),\qquad \short\to 0. \label{assump}
\ee
We use the notation of \citet{B-O} for ``$\C_i(\short)$ is much smaller than $\C_r(\short)$ as $\short$ tends to $0$'', namely that
\be
\lim\limits_{\short\to 0} \frac{\C_i(\short)}{\C_r(\short)} =0.
\ee
The purpose of such an assumption is to replace $\C\in \mathbb{C}$ by $\C_r\in \mathbb{R}$ in equations (\ref{feq}) and (\ref{heq}). Recalling that $\U(\z>0)>0$ and $\U(\z<0)<0$, if $\C_r$ is positive (negative) then there is a critical layer in the air (water), associated with a singular point in Eq.~\ref{feq} (Eq.~\ref{heq}). However, at this stage in the development, we do not know the possible values for $\C_r$. Indeed, the term $rf'(0^+)-h'(0^-)$ depends on $\C$ in a non-trivial manner so that Eq. (\ref{BC}) is not necessarily quadratic in $\C$. Physically, $\C_r$ is the phase speed of sheared interfacial waves in the reference frame of the water surface. In that frame, a wave propagating in the direction of (against) the current has a positive (negative) $\C_r$ and \citet{young-wolfe} refer to these as prograde (retrograde) modes. In the case of a constant (zero shear) current $U$, there is one prograde and one retrograde mode, which are simply the Doppler-shifted forward and backward interfacial waves. We generalize this result to the case of arbitrary shear, which we check a posteriori. Thus we assume that there are two solutions, $\C_{r+}>0$ and $\C_{r-}<0$ that correspond  to two critical levels, $\zcp>0$ and $\zcm<0$, such that
\be
\U(\zcpm) = \C_{r\pm}\label{crits}.
\ee
Therefore, the prograde (retrograde) mode undergo the Miles (rippling) instability, and hence $\C_i \neq 0$ for both modes. We stress that, provided that the values of $\C_{r\pm}$ are within the bounds of the function $\U(\z)$, critical layers actually exist. When $\C_{r\pm}$ are equal to these bounds, the system is marginally stable. 
In Appendix \ref{cutoff}, we derive a general asymptotic formula for the large wavenumber cut-off of the rippling instability. 

Hence, although $\C_{r\pm}$ are unknown, we replace $\C$ by $\C_{r+}$ in Eq. (\ref{feq}) and by $\C_{r-}$ in Eq. (\ref{heq}). We then solve these equations using boundary layer theory in \S\ref{asympsol}, where we construct uniformly valid composite solutions. However, here we need only substitute the derivatives at $\z=0^\pm$ into Eq. (\ref{BC}). In Appendix \ref{globprop}, we show that for the prograde mode, 
\begin{widetext}
\begin{align}
f'(0^+) &= -i\pi\ \frac{\U''(\zcp)}{|\U'(\zcp)|}\ e^{-2\zcp/\short} + \short\ \frac{\U''(0^+)}{2\C_{r+}(\short)} -\frac{\U''(\zcp)}{\U'(\zcp)}\ \frac{\short}{2\zcp}\ , \label{fpro}\\
\text{and}\qquad h'(0^-) &=  - \short\ \frac{\U''(0^-)}{2\C_{r+}(\short)}\ , \label{hpro}
\end{align}
whereas for the retrograde mode,
\begin{align}
f'(0^+) &=  \short\ \frac{\U''(0^+)}{2\C_{r-}(\short)}\ , \label{fretro}\\
\text{and}\qquad h'(0^-) &= i\pi\ \frac{\U''(\zcm)}{|\U'(\zcm)|} \ e^{2\zcm/\short}  - \short\ \frac{\U''(r^-)}{2\C_{r-}(\short)} +\frac{\U''(\zcm)}{\U'(\zcm)}\ \frac{\short}{2\zcm}\ . \label{hretro}
\end{align}
\end{widetext}
We can now solve Eq. (\ref{BC}).  With the advent of results (\ref{fpro}) and (\ref{hretro}), our key assumption, Eq. (\ref{assump}), is made more precise by letting 
\be
\C_\pm = \C_{r\pm}(\short) + \rm{i}\ A_\pm(\short)\ \short \ e^{-2|\zcpm|/\short}. \label{assump_precise}
\ee
Therefore, in order to have solutions with a non-zero imaginary part, we must include exponentially small terms.  When first discarding those terms, we obtain 
\begin{widetext}
\be
\frac{\s}{ \short} + \g(1-r)\short +  \big[ r \U'(0^+)-\U'(0^-)\big] \short \C_r  - (1+r + \rm{h.o.t.}) \C_r^2  =0,\qquad \short\ll1, \label{C_rEq}
\ee
\end{widetext}
where h.o.t denotes higher order terms, such as terms of order $\short/\C_r$ and $\short/\zc$.  

We seek solutions of Eq. (\ref{C_rEq}) as a series in powers of $\short$. The nature of the leading order term depends on the value of the dimensionless surface tension $\s$. When gravity is the only restoring force, we find 
\begin{widetext}
\be
\C_{r\pm}^{\rm{grav}}= \pm \sqrt{\frac{1-r}{1+r}\ \g\short}\ + \frac{r\U'(0^+)-\U'(0^-)}{2(1+r)}\ \short\ \pm \frac{[r\U'(0^+)-\U'(0^-)]^2}{8\sqrt{\g(1-r)}}  \bigg[\frac{\short}{1+r}\bigg] ^{\frac{3}{2}} + \rm{h.o.t}. \label{grav}
\ee
When capillary forces are included, we instead obtain 
\be
\C_{r\pm}^{\rm{cap-grav}}= \pm \sqrt{\frac{\s}{(1+r)\short}} + \frac{r\U'(0^+)-\U'(0^-)}{2(1+r)}\ \short\ \pm \frac{\g(1-r)}{2\sqrt{\s(1+r)}}\ \short^{\frac{3}{2}} + \rm{h.o.t}. \label{cap-grav}
\ee
\end{widetext}
Hence, apart from the effect of shear, for short waves  gravity acts as a high order correction to the effect of surface tension. Note that the third term in Eq. (\ref{cap-grav}) can be derived by expanding the dispersion relation of interfacial capillary-gravity waves, 
\be
\C(\short)= \pm  \sqrt{\frac{\s}{(1+r)\short} + \frac{1-r}{1+r}\ \g\short} \ . \label{free}
\ee
Armed with $\C_{r\pm}$, we can use assumption (\ref{assump_precise}), and Eqs. (\ref{fpro}) and (\ref{hretro}), in Eq. (\ref{BC}).  We collect terms of order $\short \ e^{-2|\zcpm|/\short}$ to find the amplitudes $A_\pm$, and infer the growth rates of the prograde (+) and retrograde (-) modes as
\begin{align}
\Im\{\kd\C_+\} &= - \C_{r+}\ \frac{\pi}{2}\ \frac{r}{1+r}\ \frac{\U''(\zcp)}{|\U'(\zcp)|}\ e^{-2\kd\zcp}, \label{gr+} \\
\text{and}\qquad\Im\{\kd\C_-\} &= - \C_{r-}\ \frac{\pi}{2}\ \frac{1}{1+r}\ \frac{\U''(\zcm)}{|\U'(\zcm)|}\ e^{2\kd\zcm}, \label{gr-}
\end{align}
where $\kd=1/\short$. We emphasize several aspects of the present results. Firstly, equation (\ref{gr+}) for the prograde mode is a generalization of the growth rate obtained by Miles in an Appendix of \citet{morland-saffman}. That result was obtained from short wavelength asymptotic analysis of the exact solution of the Rayleigh equation for an exponential wind profile. Secondly, equation (\ref{gr-}) for the retrograde mode is a generalization of the short wavelength limit of the growth rate of the rippling instability found by \citet{young-wolfe} for an exponential wind-induced current. \textcolor{black}{Here, we have included the effect of the upper fluid on the rippling instability, showing the weakness of the effect for the air-water system because $r=O(10^{-3})$, consistent with the $r=0$ limit of \citet{young-wolfe}. Finally, we find that the results of \citet{shrira} are a special case of our own when $r=0$, but stress that his small parameter was the smallness of the deviation of the wave motion from potential flow, rather than the inverse wavenumber.}
\begin{figure*}[htbp!]
        (a)\includegraphics[trim = 0 0 0 0, clip, width = 0.33\textwidth]{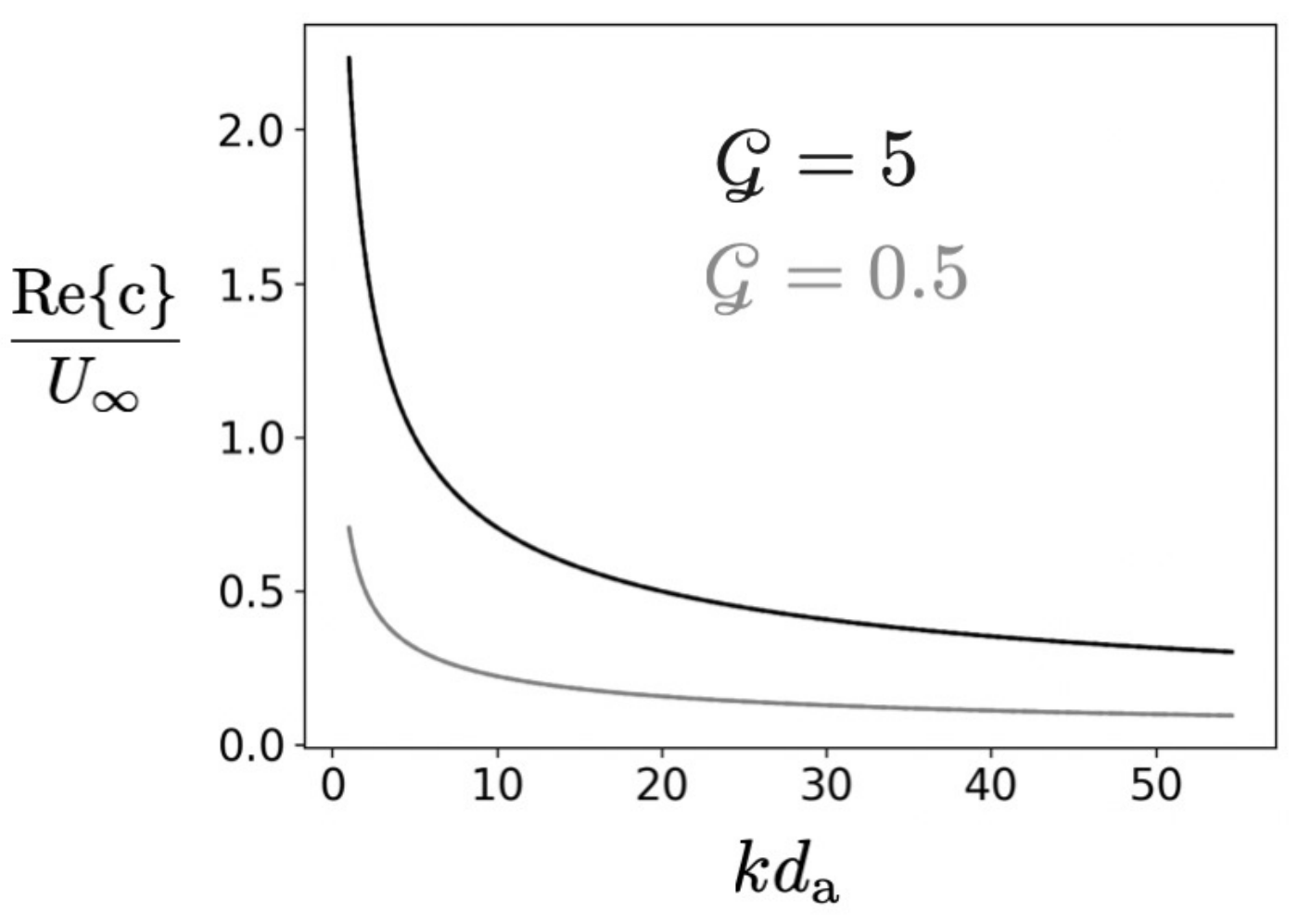}	  	
        (b)\includegraphics[trim = 0 0 0 0, clip, width = 0.28\textwidth]{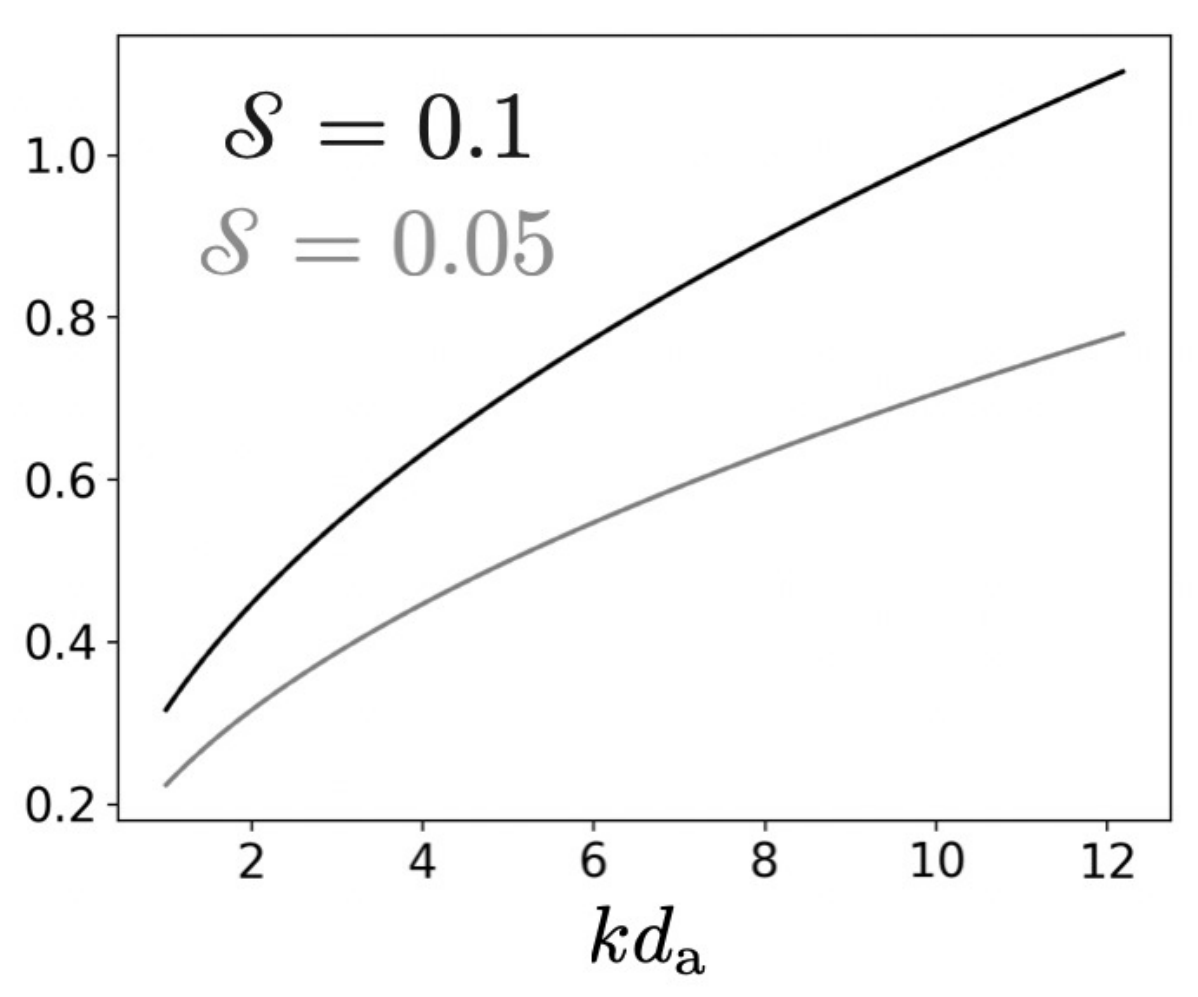}
        (c)\includegraphics[trim = 0 0 0 0, clip, width = 0.29\textwidth]{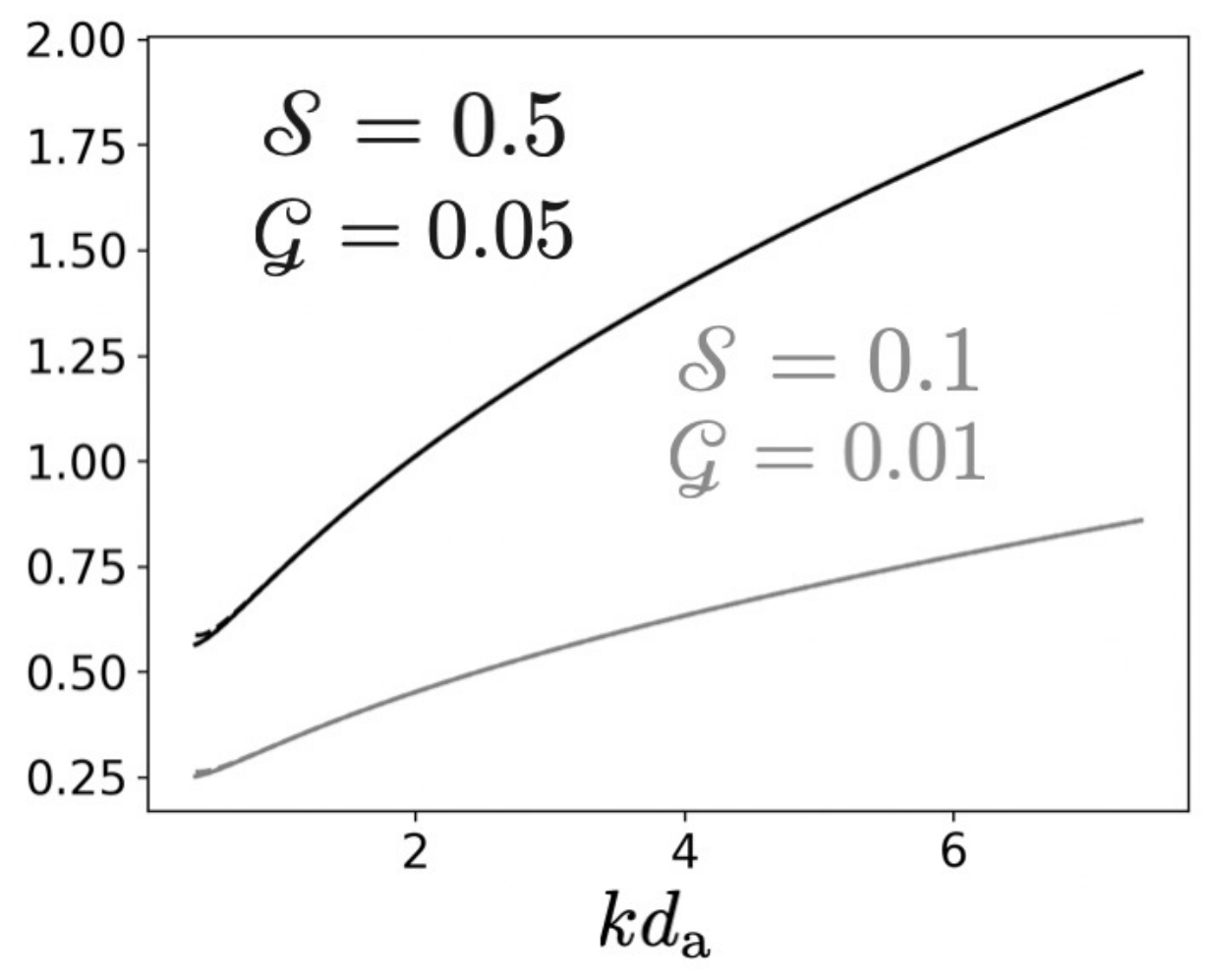}	  	
        (d)\includegraphics[trim = 0 0 0 0, clip, width = 0.35\textwidth]{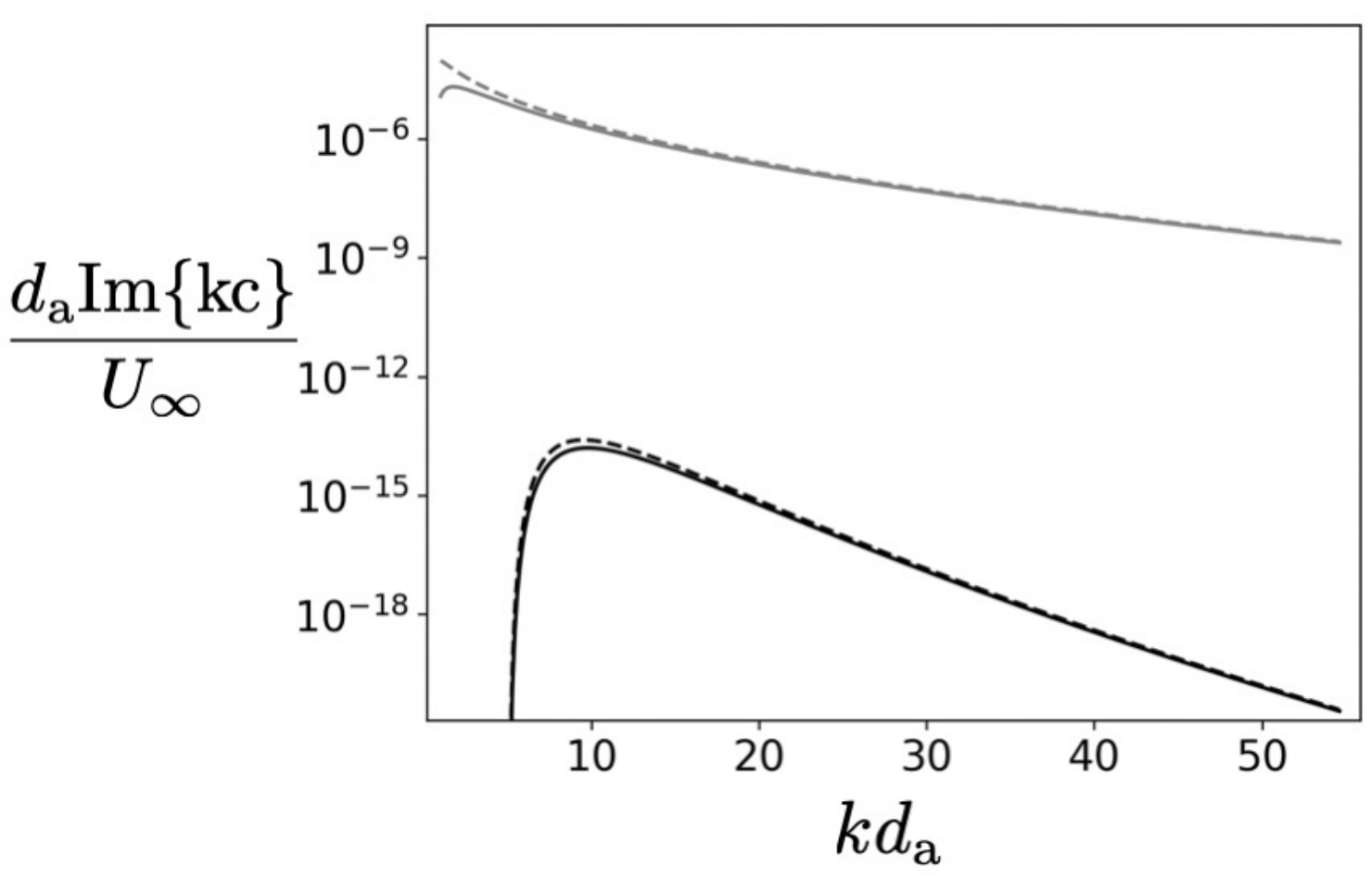}
        (e)\includegraphics[trim = 0 0 0 0, clip, width = 0.28\textwidth]{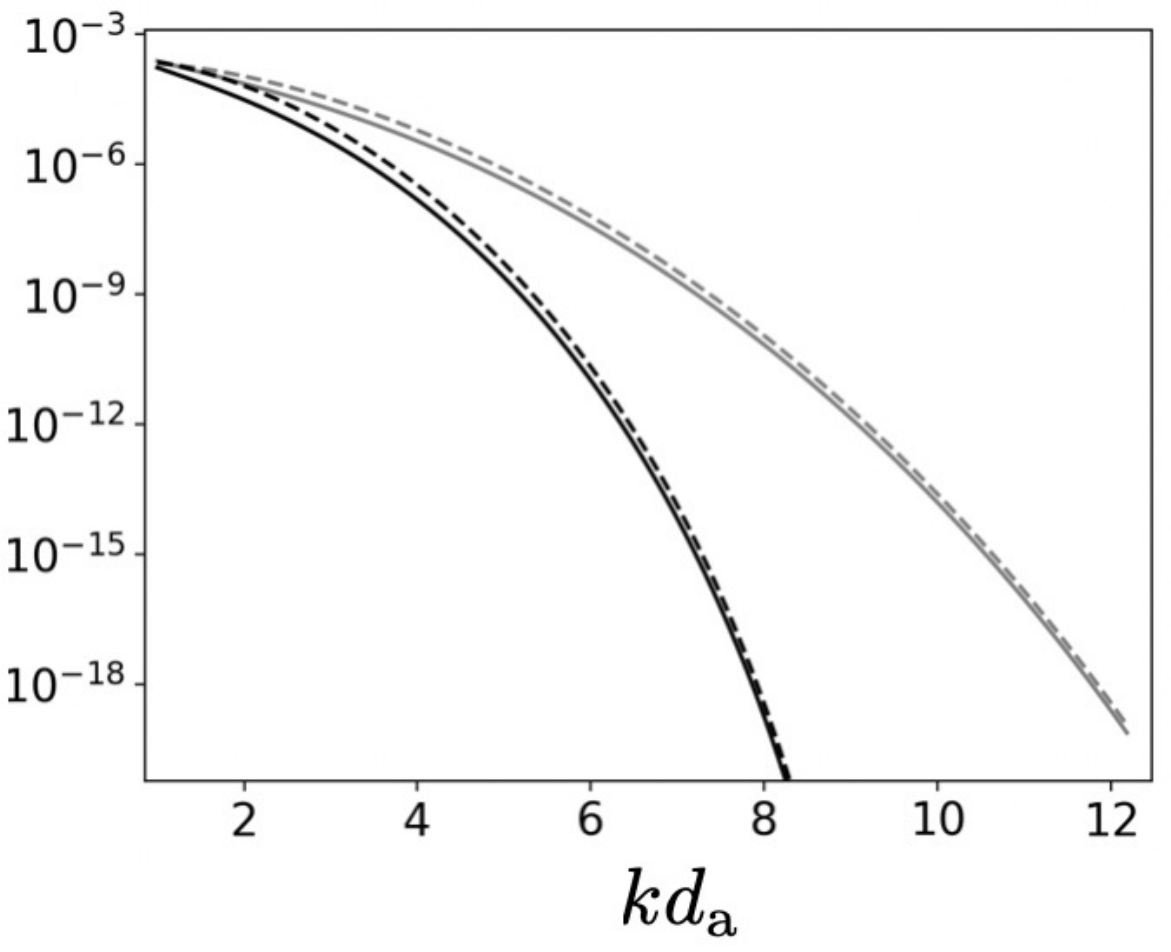}	  	
        (f)\includegraphics[trim = 0 0 0 0, clip, width = 0.28\textwidth]{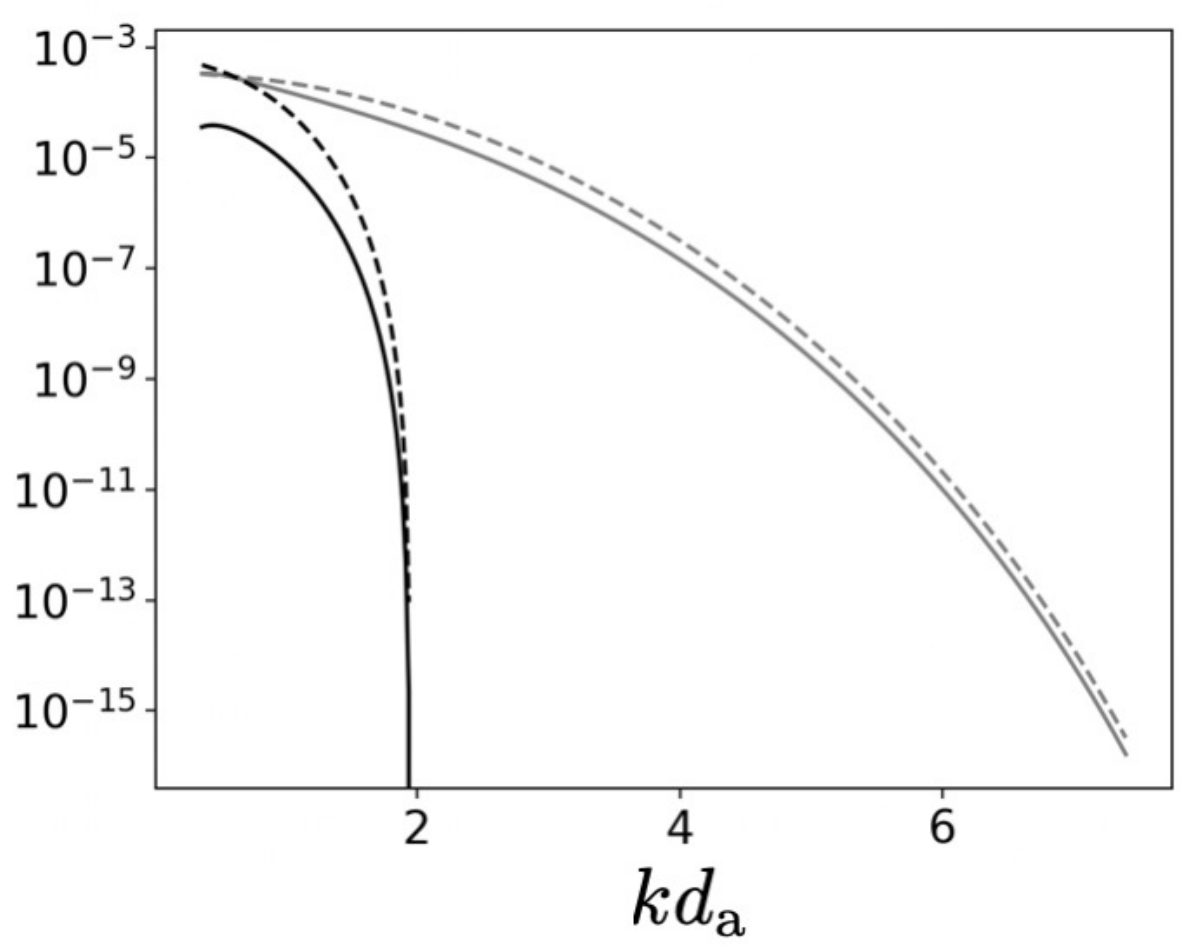}
  \caption{Comparison of the short wavelength asymptotic results (dashed lines) for the prograde mode with the exact solution (solid lines) in the case of an exponential wind profile (no current), $U(z)= U_\infty\ (1-e^{-z/\da})$, a density ratio $r=0.001$, and different values (grey and black) of the dimensionless gravity, $\g=g\da/U_\infty^2$, and surface tension, $\s=\sigma/(\dwat U_\infty^2\da)$.
The phase speed as a function of the wavenumber is shown for gravity~(a), capillary (b), and capillary-gravity waves (c); the exact and the asymptotic solutions are indistinguishable. The growth rate as a function of the wavenumber for gravity (d), capillary (e), and capillary-gravity waves~(f).}
\label{smallratio}
\end{figure*}
\begin{figure*}[htbp!]
        (a)\includegraphics[trim = 0 0 0 0, clip, width = 0.33\textwidth]{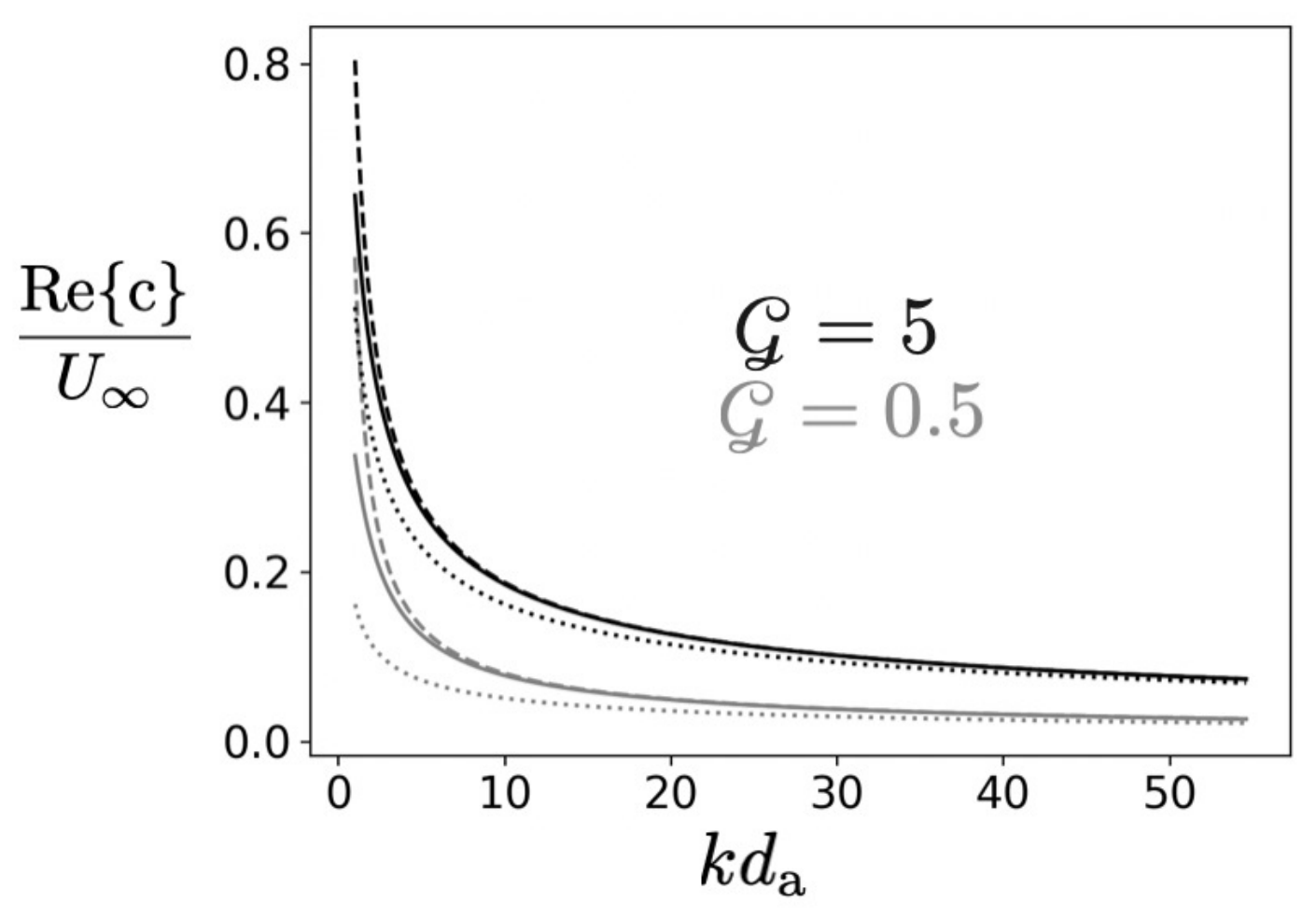}	  	
        (b)\includegraphics[trim = 0 0 0 0, clip, width = 0.28\textwidth]{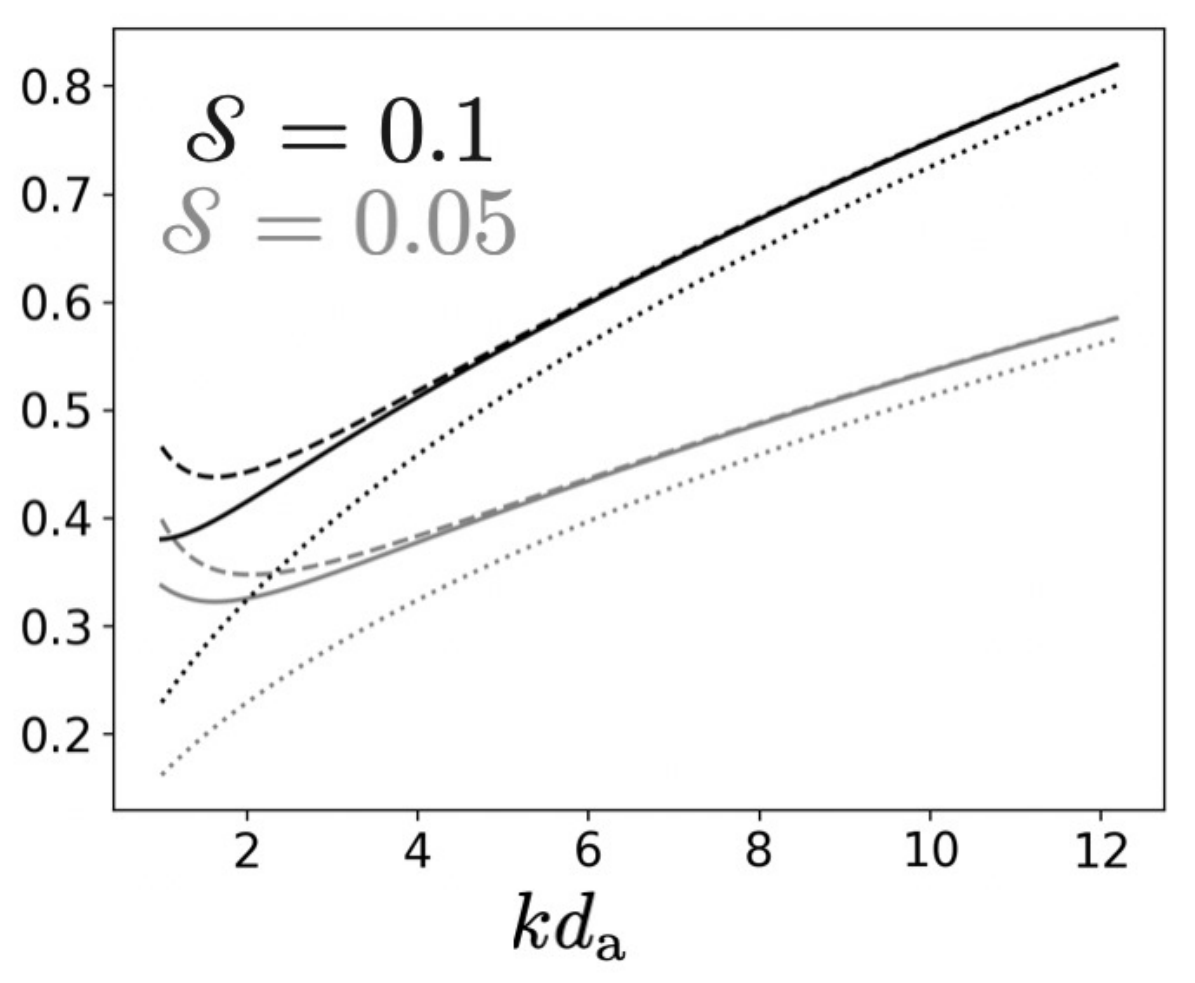}
        (c)\includegraphics[trim = 0 0 0 0, clip, width = 0.28\textwidth]{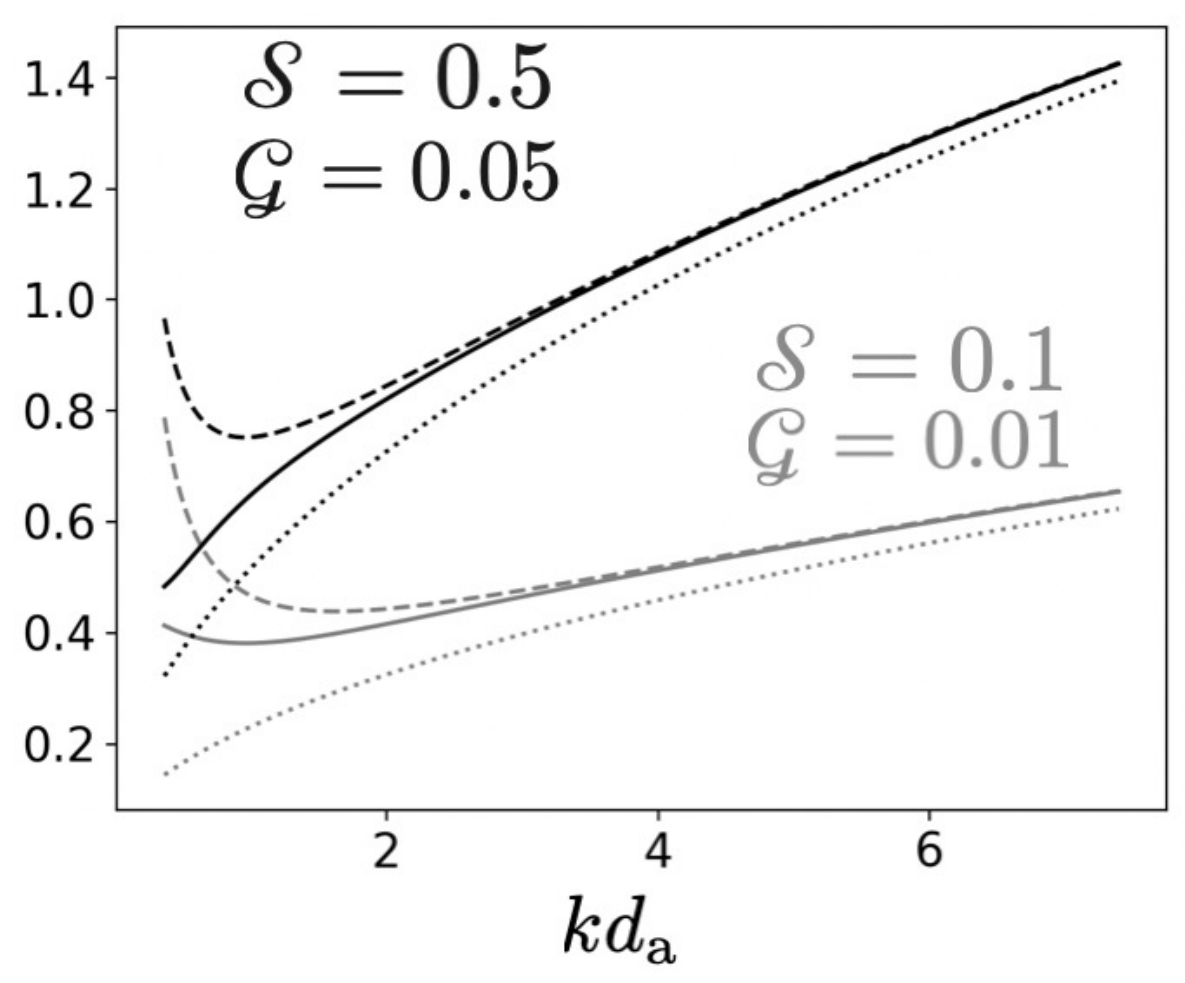}	  	
        (d)\includegraphics[trim = 0 0 0 0, clip, width = 0.35\textwidth]{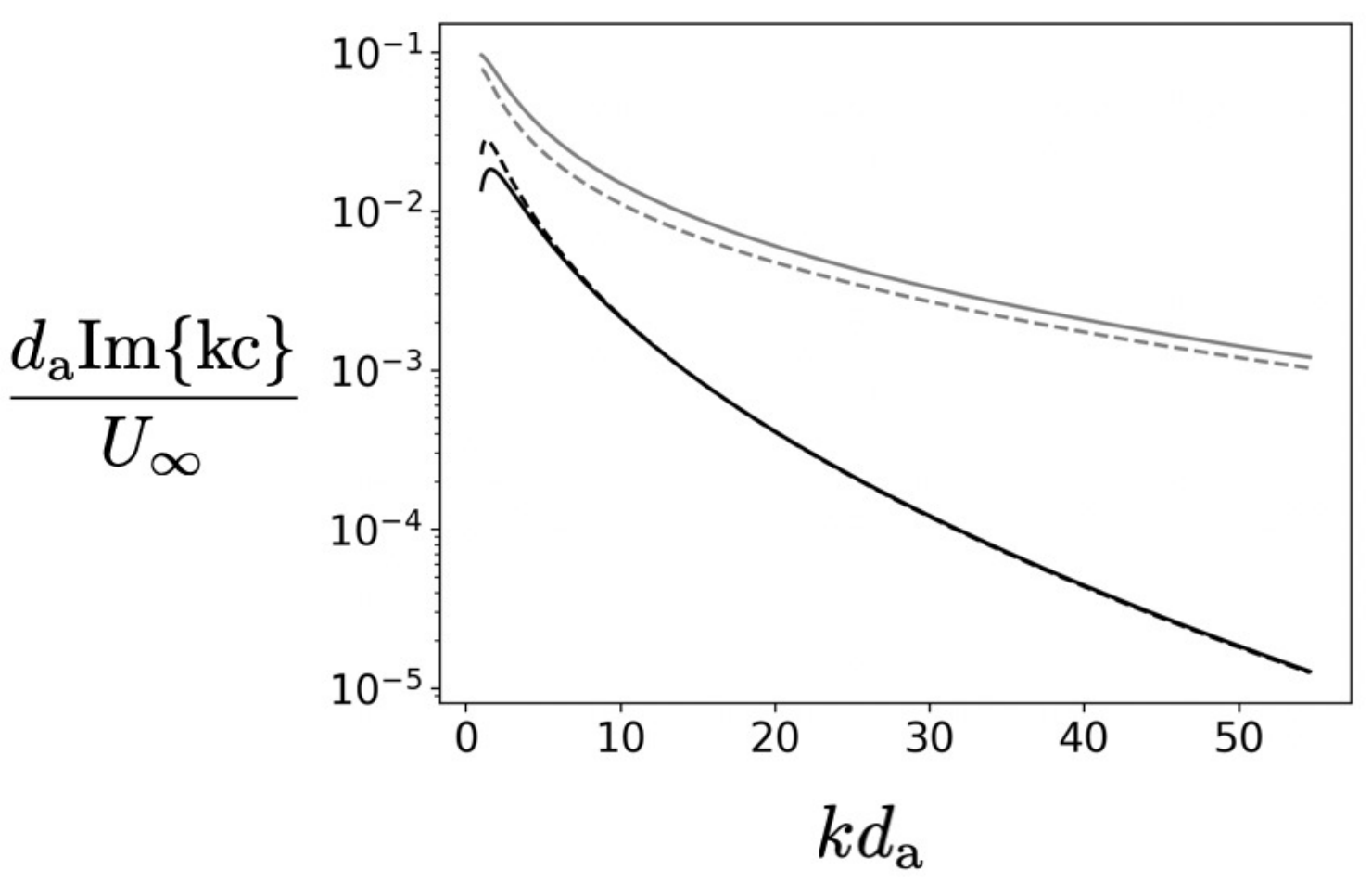}
        (e)\includegraphics[trim = 0 0 0 0, clip, width = 0.28\textwidth]{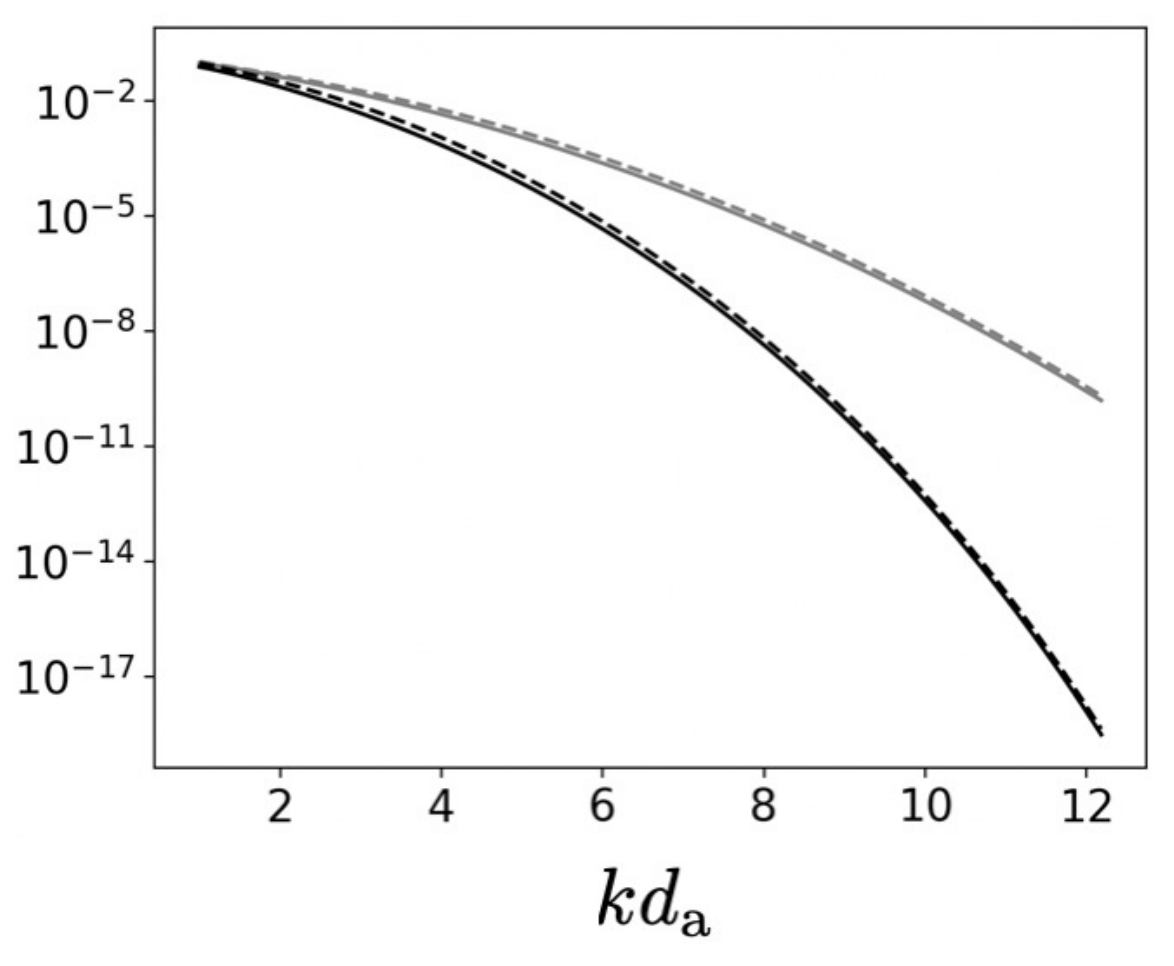}	  	
        (f)\includegraphics[trim = 0 0 0 0, clip, width = 0.28\textwidth]{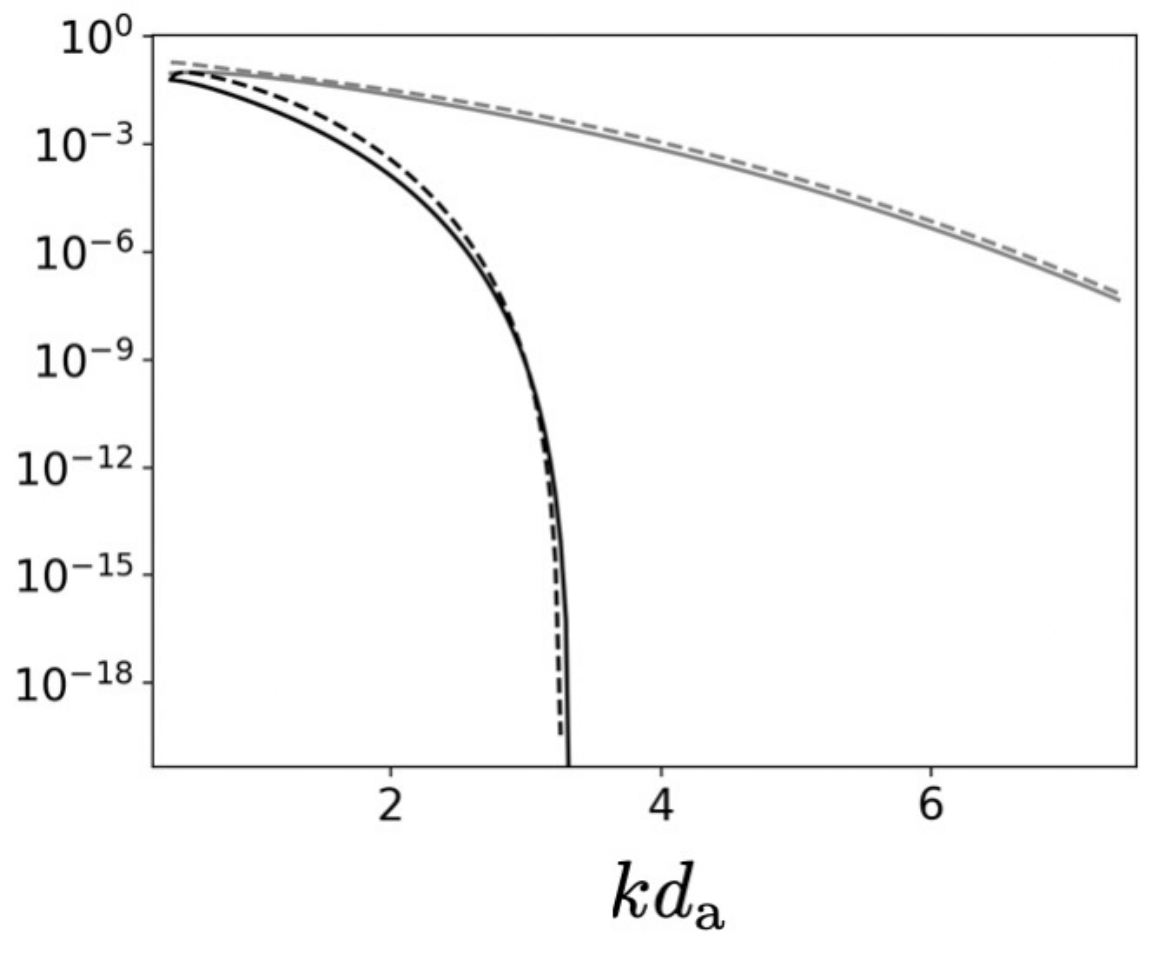}
  \caption{Same as Fig.~\ref{smallratio} but with a density ratio of $r=0.9$. Whereas in Figs.~\ref{smallratio}(a)-(c) the asymptotic results (dashed lines) and the exact solution (solid lines) are indistinguishable, such is not the case here, where we also see the phase speed of free surface waves (dotted lines).}
\label{largeratio}
\end{figure*}

\section{Interpretation} \label{discussion}

\begin{figure*}[htbp!]
        (a)\includegraphics[trim = 0 0 0 0, clip, width = 0.33\textwidth]{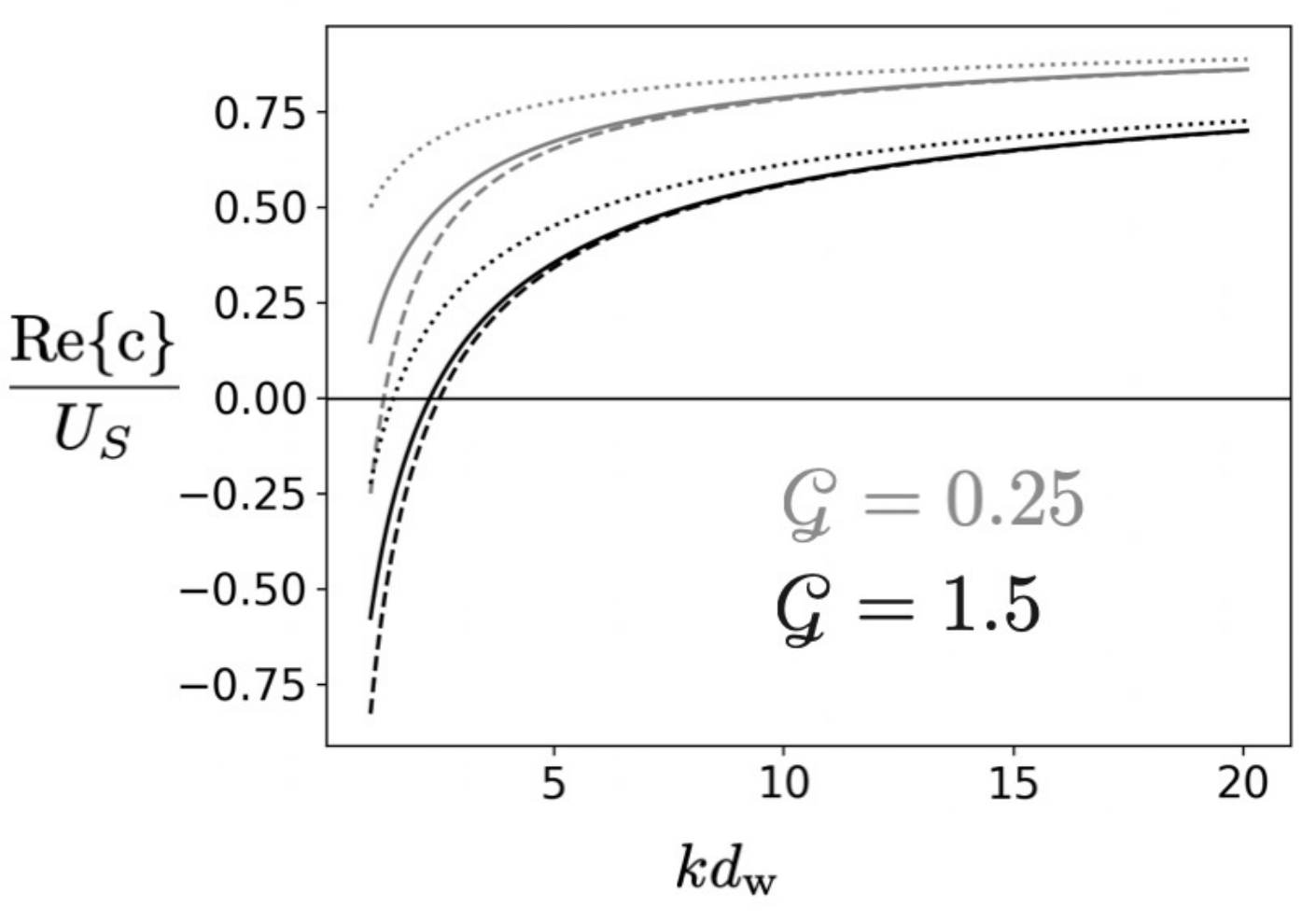}	  	
        (b)\includegraphics[trim = 0 0 0 0, clip, width = 0.29\textwidth]{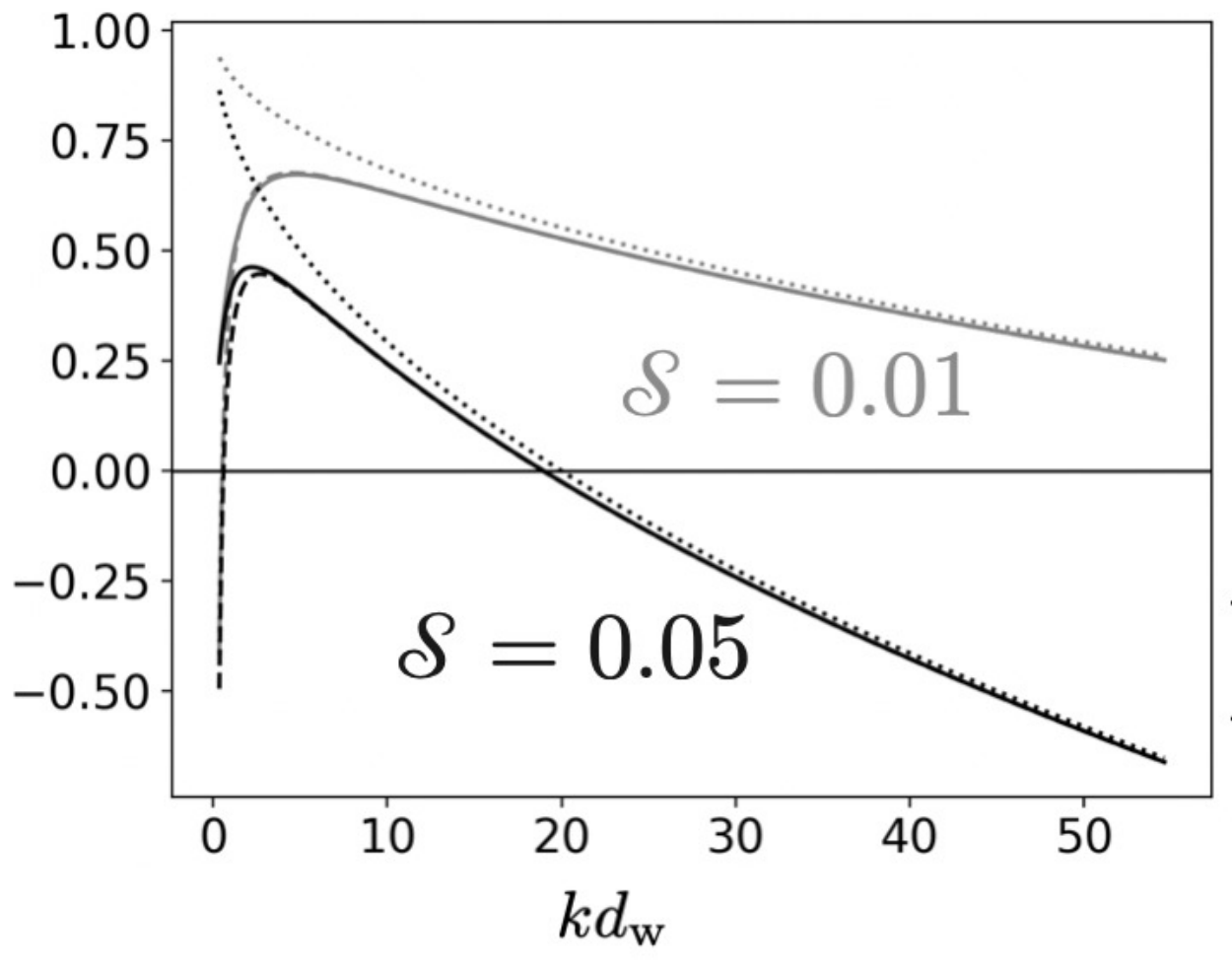}
        (c)\includegraphics[trim = 0 0 0 0, clip, width = 0.29\textwidth]{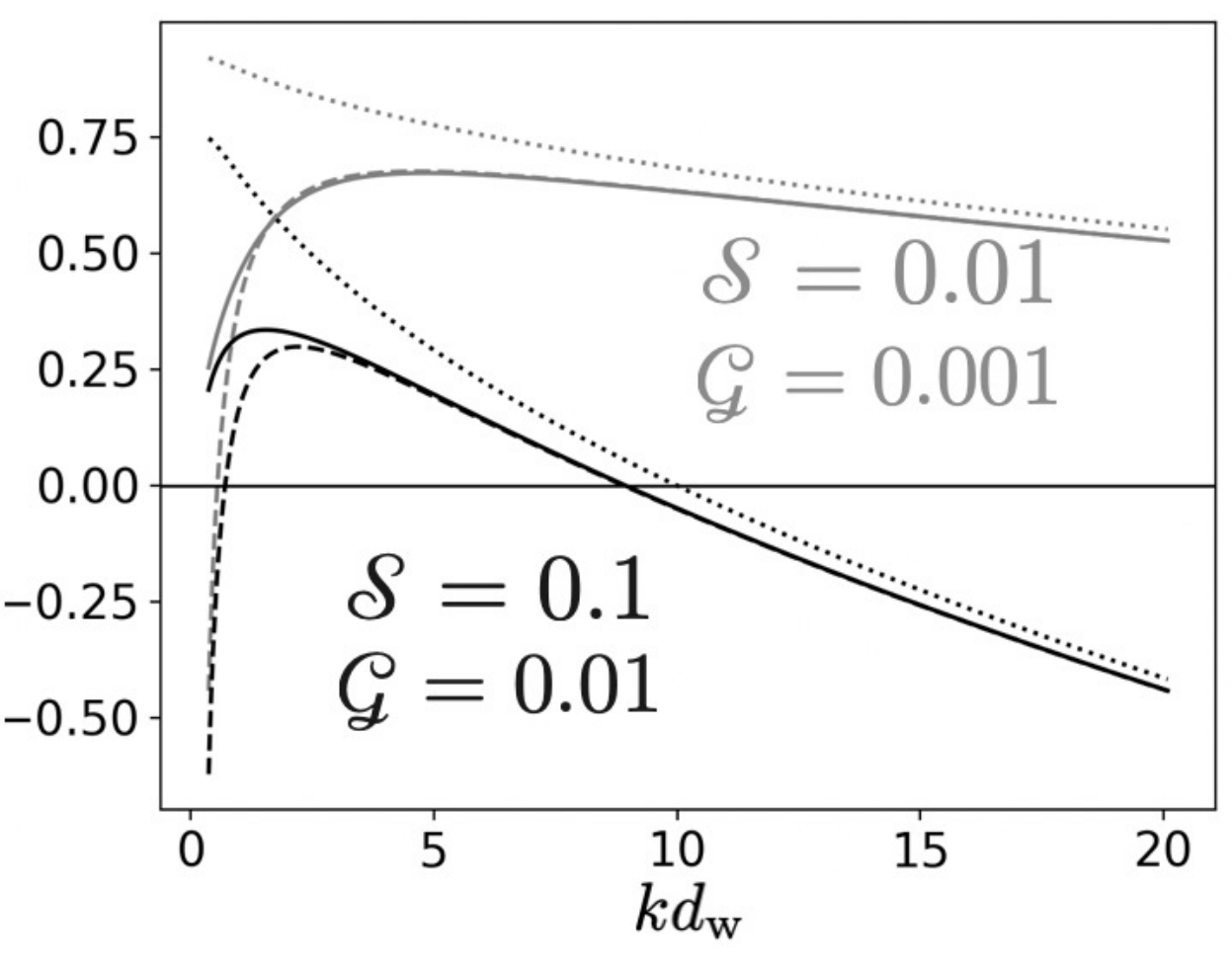}	  	
        (d)\includegraphics[trim = 0 0 0 0, clip, width = 0.34\textwidth]{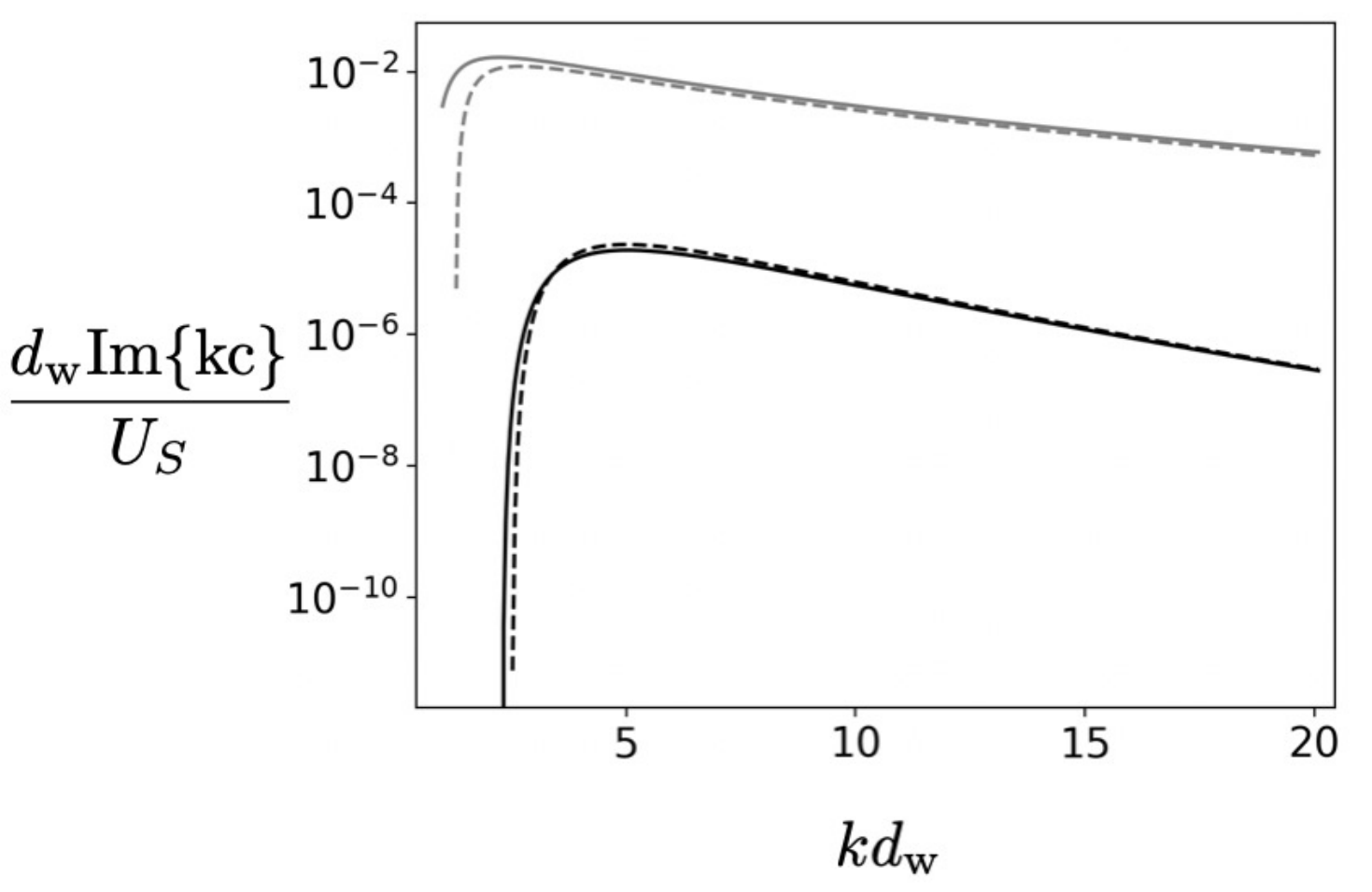}
        (e)\includegraphics[trim = 0 0 0 0, clip, width = 0.28\textwidth]{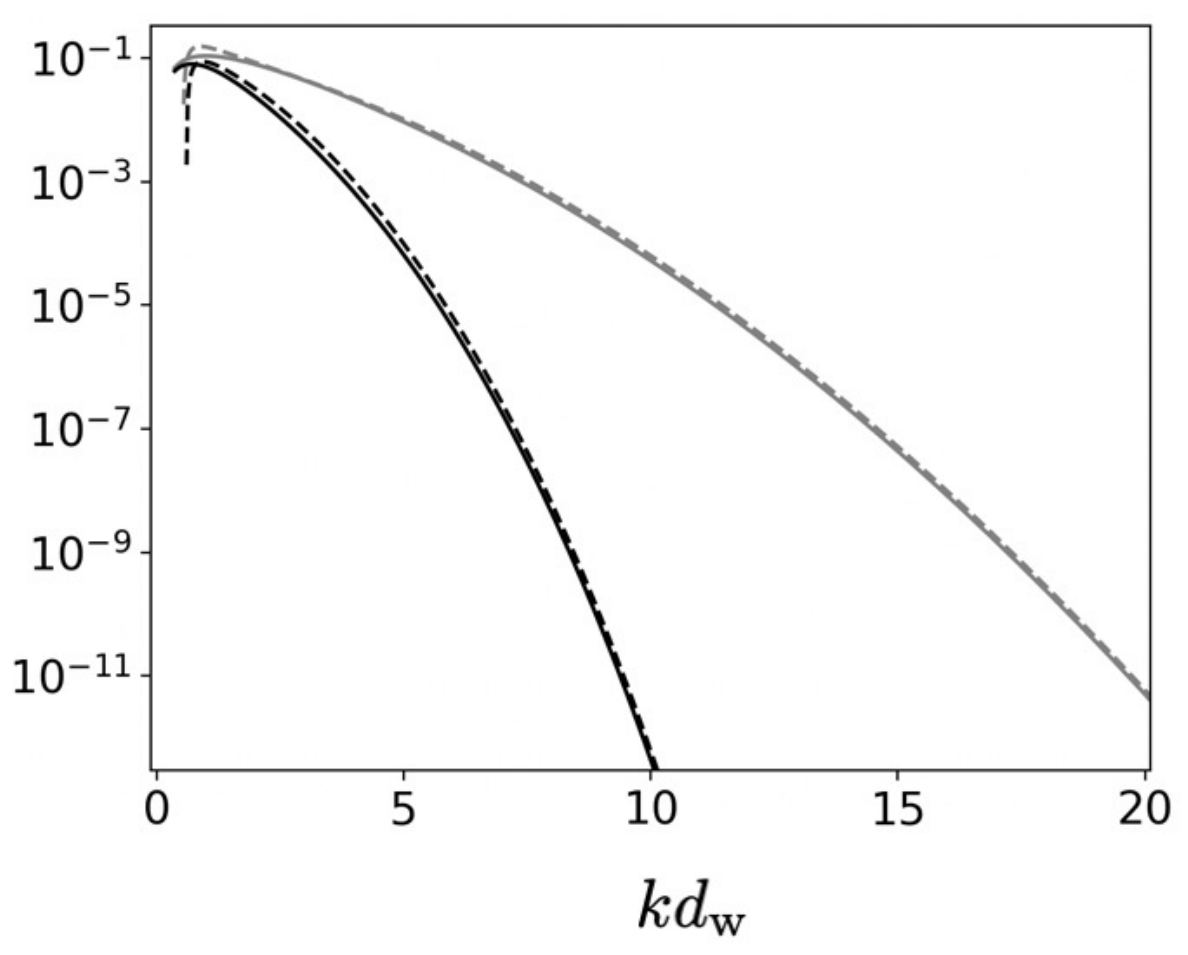}	  	
        (f)\includegraphics[trim = 0 0 0 0, clip, width = 0.28\textwidth]{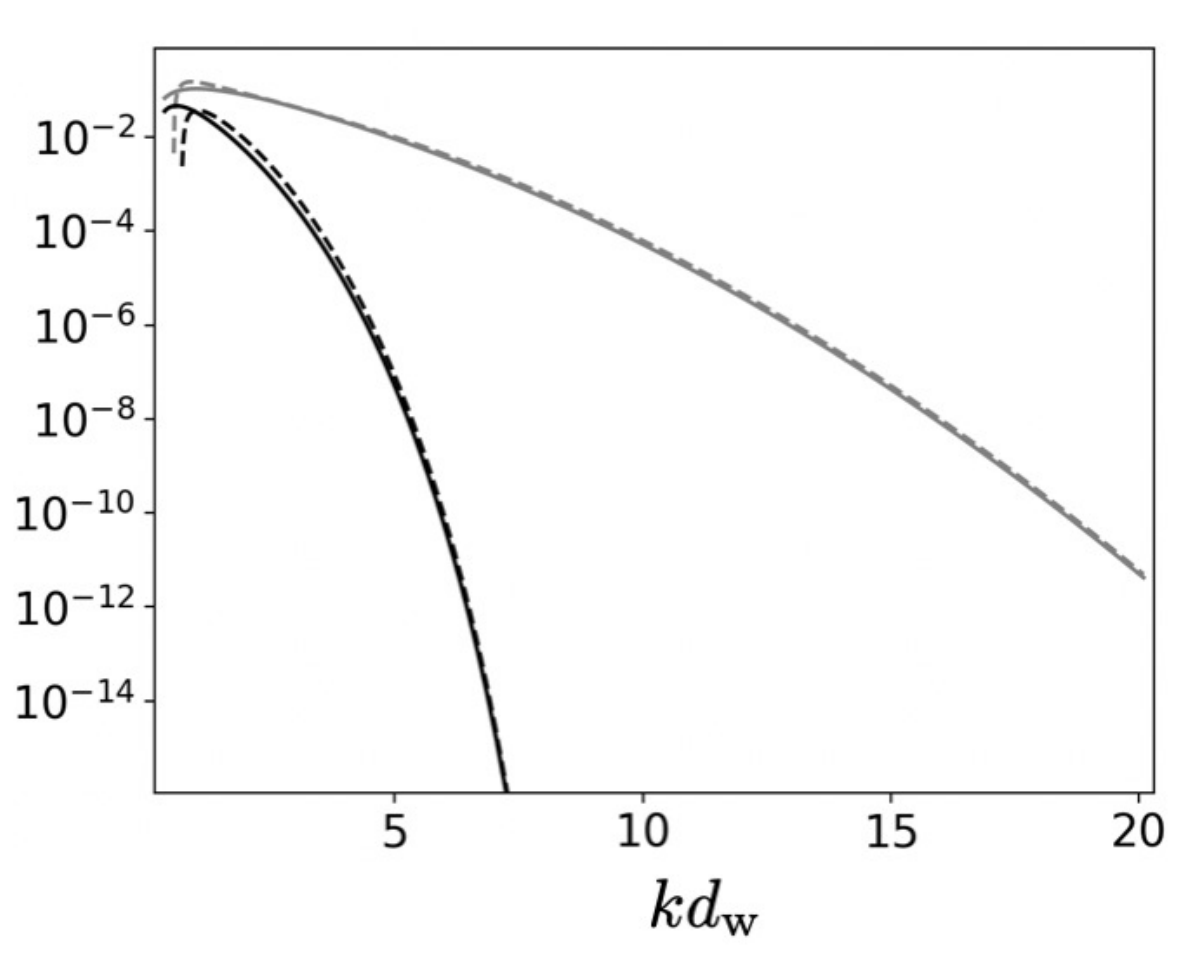}
  \caption{Comparison of the short wavelength asymptotic results (dashed lines) for the retrograde mode with the exact solution (solid lines) in the case of an exponential current (no air), $U(z)= \Us\ e^{z/\dw}$, 
  and different values (grey and black) of the dimensionless gravity, $\g=g\dw/\Us^2$, and surface tension, $\s=\sigma/(\dwat \Us^2\dw)$.  The dotted lines depict the corresponding phase speed of Doppler-shifted surface waves. The growth rate as a function of the wavenumber for gravity (d), capillary (e), and capillary-gravity waves~(f).}
\label{rippling}
\end{figure*}
\begin{figure*}[htbp!]
  \centerline{\includegraphics[scale=0.22]{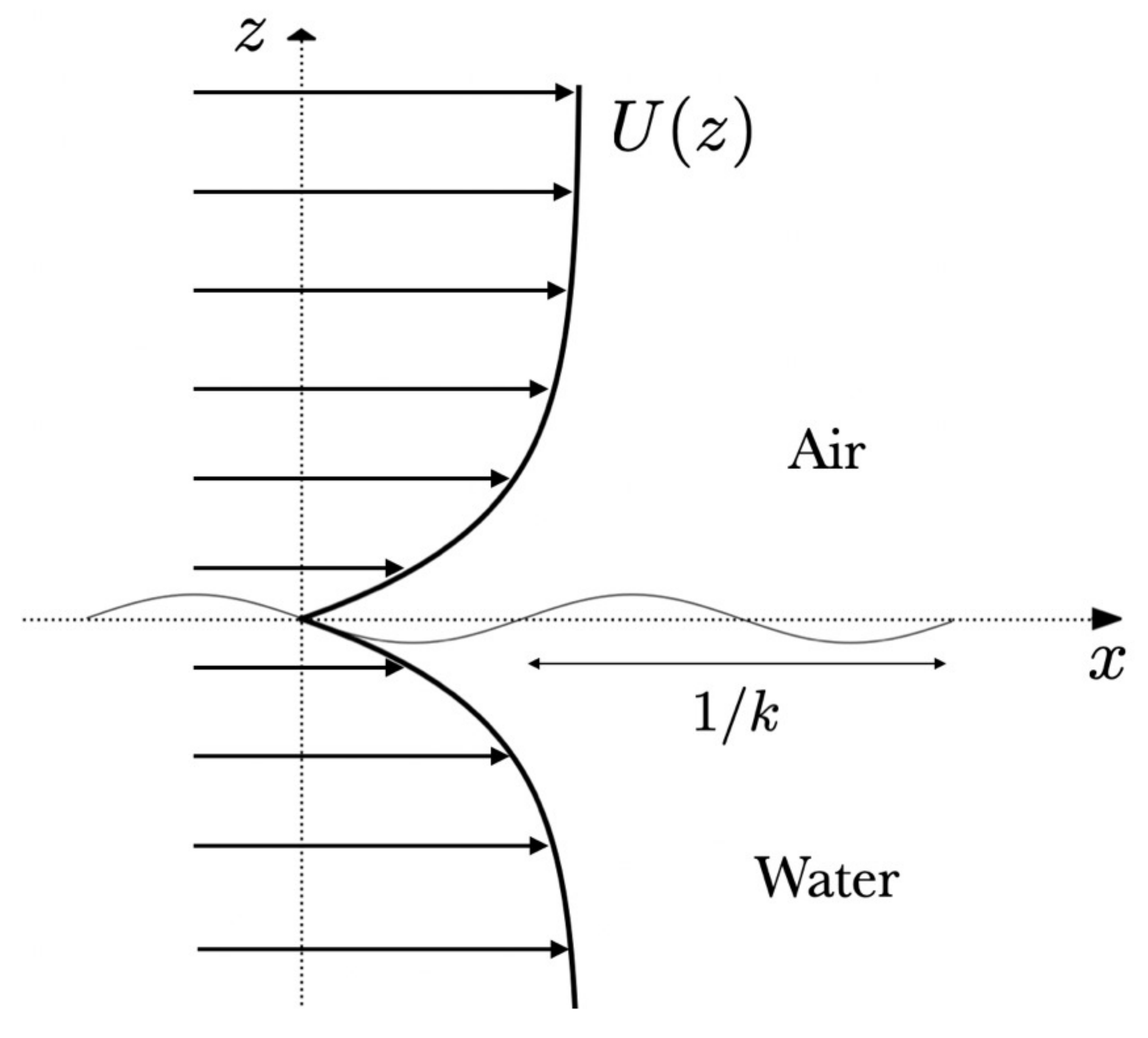}}
  \caption{Schematic of a wind blowing in the air and a current in the water. }
\label{externalcurrent}
\end{figure*}
The behavior of short waves in presence of a wind and a wind-induced current depends on the profile of the latter solely through the derivatives at $\z=0$ and at the critical levels, $\z=\zcpm$. In consequence, the results are qualitatively the same for the two profiles introduced in \S \ref{profiles}. We note that, as shown by \citet{young-wolfe},  the eigenvalue problem can be solved exactly in terms of hypergeometric functions in the case of double exponential profiles, whereas exact analytical solutions for double log profiles are unknown.

In figures \ref{smallratio} and \ref{largeratio} we show the phase speed and the growth rate of the prograde mode undergoing the Miles instability.  We compare the results of our short wavelength asymptotic analysis with the exact solution for an exponential wind profile, in the case of gravity (a and d), capillary (b and e), and capillary-gravity waves (c and f), for different values of the density ratio $r$ and the control parameters $\g$ and $\s$. Figure \ref{rippling} shows a similar comparison for the retrograde mode undergoing the rippling instability in the absence of air, that is $r=0$. The asymptotic results are in accord with the exact results. 

The phase speed of Doppler-shifted interfacial waves in presence of a constant, zero shear, current is 
\be
c(k)=\Us \pm \sqrt{\frac{1-r}{1+r}\ \frac{g}{k}+\frac{\sigma}{\dwat(1+r)}\ k}\ , \label{free_dim}
\ee
which is the dimensional form of Eq. (\ref{free}), where the plus (minus) sign corresponds to the prograde (retrograde) mode. Equations (\ref{grav}) and (\ref{cap-grav}) predict that the non-uniformity of the current increases the phase speed by $r U'(0^+)/2k$ and reduces it by $U'(0^-)/2k$ respectively. Indeed, Figures \ref{largeratio}(a)-(c) show that when $r$ is close to $1$, and $\kd$ approaches unity, the phase speed of sheared waves is significantly larger than the phase speed of free surface waves. Moreover, Figures \ref{rippling}(a)-(c)  show the predicted reduction of the retrograde mode in the absence of air when $\kd$ is of order unity.

Generally, for small values of $r$, the effect of the wind on dispersion is insignificant, as shown by the phase speed of free surface in Figures \ref{smallratio}(a)-(c). However, for large values of $r$, the wind has a major influence. For instance, as shown in Figure \ref{largeratio}(b), the wind is responsible for a minimum phase speed of capillary waves. Moreover, as $r$ increases so does the growth rate of the prograde mode. Thus, we interpret the density ratio as a wind-wave coupling constant.

Finally, for both modes the growth rates increase when $\g$ and $\s$ are small, which is the case for large velocity scales (c.f., Figs~\ref{smallratio} and \ref{rippling}). Thus, consistent with physical intuition, a small perturbation grows faster in the presence of strong winds and/or currents. 
\section{Prograde instability due to critical layers in both air and water} \label{doublecrit}
We now explore the case of a wind blowing in the air when there is also a current in the water, as shown in Figure~\ref{externalcurrent}. Thus we have $\U\big(\z\lessgtr0\big)>0$, $\U'(\z>0)>0$, and $\U'(\z<0)<0$, and again assume an infinite depth and proceed as in \S \ref{shortwave}. The key difference is that the prograde mode has a critical layer in both the air and the water, and the retrograde mode has no critical layer.  Therefore, there are levels $\zcpm$ such that
\be
\U(\zcp) = \C_{r+}\qquad\text{and}\qquad \U(\zcm) = \C_{r+},
\ee
and there is no value of $\z$ such that $\C_{r-}<0$ is equal to $\U(\z)$. This implies that
\begin{widetext}
\begin{align}
f'(0^+) &= -i\pi\ \frac{\U''(\zcp)}{|\U'(\zcp)|} \ e^{-2\zcp/\short}  + \short\ \frac{\U''(0^+)}{2\C_{r+}(\short)}  -\frac{\U''(\zcp)}{\U'(\zcp)}\ \frac{\short}{2\zcp}\ ,  \\
 \text{and}\qquad h'(0^-) &= i\pi\  \frac{\U''(\zcm)}{|\U'(\zcm)|}\ e^{2\zcm/\short}  - \short\ \frac{\U''(0^-)}{2\C_{r+}(\short)} + \frac{\U''(\zcm)}{\U'(\zcm)}\ \frac{\short}{2\zcm}\ ,
\end{align}
\end{widetext}
for the prograde mode, while 
\begin{align}
f'(0^+) &=  \short\ \frac{\U''(0^+)}{2\C_{r-}(\short)}\ , \\
\text{and}\qquad h'(0^-) &= - \short\ \frac{\U''(0^-)}{2\C_{r-}(\short)}\ , 
\end{align}
for the retrograde mode. Hence, the retrograde mode is neutral but the prograde mode can undergo both the Miles and rippling instabilities, and has a growth rate of
\begin{widetext}
\be
\Im\{\kd\C_+\} =  - \frac{\pi }{1+r}\ \frac{\C_{r+}}{2}\bigg[ r\ \frac{\U''(\zcp)}{|\U'(\zcp)|}\  e^{-2\kd\zcp}+\frac{\U''(\zcm)}{|\U'(\zcm)|}\  e^{2\kd\zcm} \bigg]. 
\ee
\end{widetext}
The real parts $\C_{r\pm}$ are still given by equations (\ref{grav}) and (\ref{cap-grav}). Such an enhancement of the Miles instability by the rippling instability may be difficult to observe in geophysical flows. However, it could be examined in a controlled laboratory setting through a refinements of the viscosity-stratified approach of \citet{Charles:1965} or the two-layer Couette flow approach of \citet{two-layer}, as well as in the context of Holmboe wave experiments \citep[e.g.,][]{Carpenter:2010}.

\section{Asymptotic solution of the Rayleigh equation for short waves} \label{asympsol}

\begin{figure}
  \centerline{\includegraphics[scale=0.2]{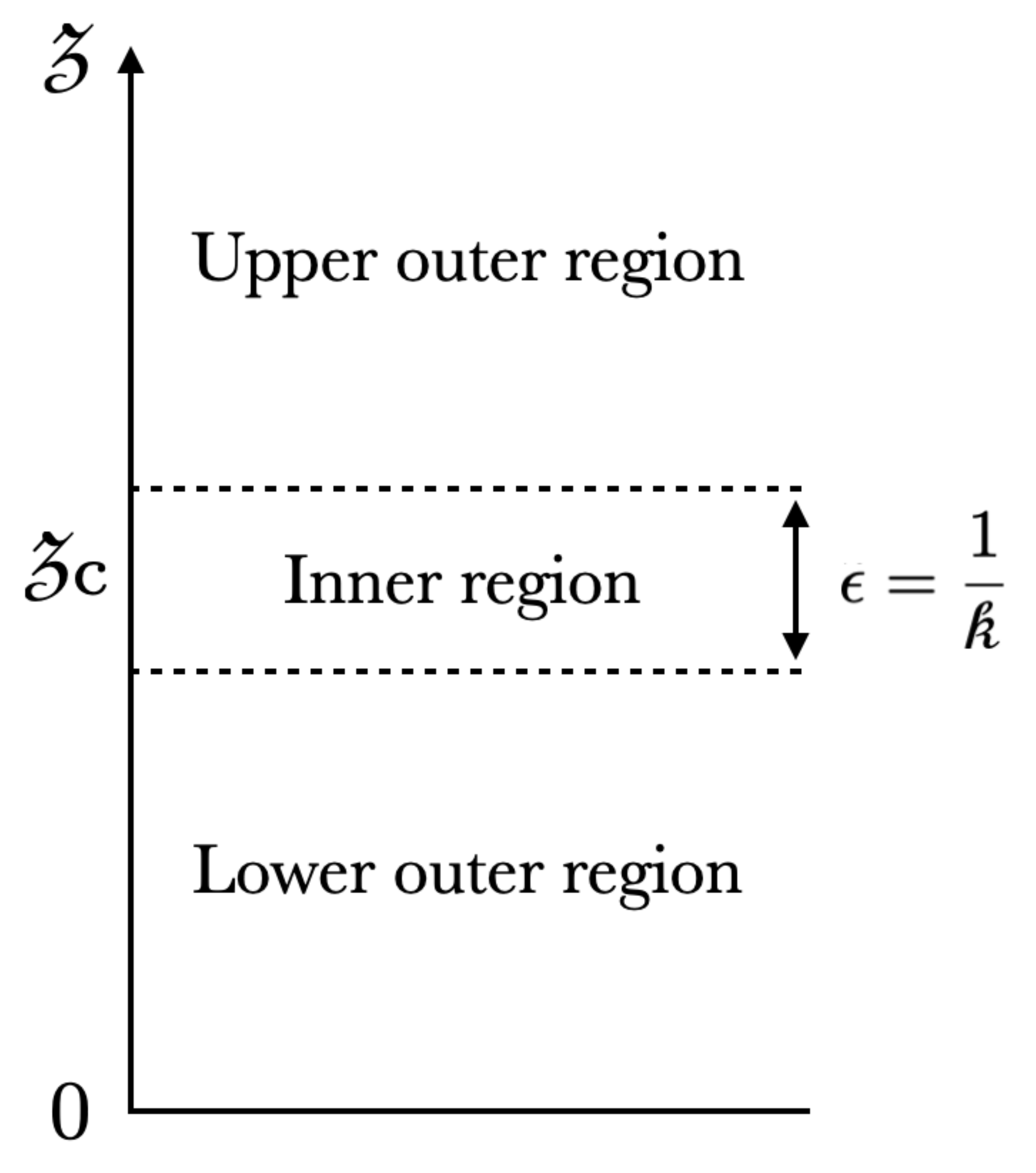}}
  \caption{Schematic of the domain of analysis of the Rayleigh equation for short waves; $\kd\gg 1$. A boundary layer of thickness $\short=1/\kd$ emerges from the singularity at the critical level, $\z=\zcd$.}
\label{BL}
\end{figure}
Here we solve equation (\ref{feq}) for $\C=\C_{r+}$ and note that a similar procedure is applicable to equation (\ref{heq}) when $\C=\C_{r-}$. For simplicity, we rewrite equation (\ref{feq}) 
\be
\short f''(\z)- 2 f'(\z)- \short\ \frac{\U''(\z)} {  \U(\z)-\C_r }\ f(\z)=0, \qquad  f(0)=1, \nonumber
\ee
and we drop the subscript $+$ for the rest of this section.
There is a regular singularity at $\z=\zcd$ such that 
\be
\U(\zcd) = \C_r. \nonumber
\ee
Because the small parameter $\short$ multiplies the highest order derivative in equation~(\ref{feq}), we expect the solution to have a boundary layer somewhere in the domain, but do not know its location a priori. However, guidance is provided by the presence of the singularity. The Frobenius exponents are $0$ and $1$, so that the solution of equation (\ref{feq}) is finite at $\z=\zcd$ whereas its derivative has a logarithmic divergence \citep{drazin-reid}. Therefore, we assume that an internal boundary layer emerges from the singularity. We assume that the point $\z=\zcd(\short)$ is well separated from the lower boundary, $\z = 0$, as $\short$ goes to zero. 
We check this a posteriori once the dependence of $\C_r$ on $\short$ is known. In consequence, $\C_r$ can be treated as a constant in the following analysis. 

The boundary layer is an inner region where the solution of equation (\ref{feq}) changes rapidly. We define two outer regions, where the solution changes slowly. One spans $\z=0$ to the inner region; the other spans the inner region to the far field. We call them `lower outer region' and `upper outer region', respectively (see Figure \ref{BL}). 

\subsection{Outer solutions}

Following \citet{miles62}, we seek outer solutions as a power series in $\short$ as
\be
\f(\z)=f_0(\z) + \short\ f_1(\z) + O\big(\short^2\big),\qquad\short\to0^+\label{out}, 
\ee
and find that
\refstepcounter{equation}
$$
f_0 = \text{constant}\quad\text{and}\quad f_1(\z) = -\frac{f_0}{2} \int^{\z}_{\z_1} d\tilde{z}\   \frac{\U''(\tilde{z})} {  \U(\tilde{z})-\C_r }\ .\label{outershort}
  \eqno{(\theequation{\mathit{a},\mathit{b}})}
$$
The constant $f_0$ and the lower limit of the integral $\z_1$ are not a priori the same for the two outer solutions. In particular, they cannot be determined by the boundary conditions at $z=0$ and infinity, requiring us to find an inner solution.

\subsection{Inner solution}

Within the boundary layer, we introduce the stretched coordinate $Z\equiv(\z-\zcd)/\delta$, where $0<\delta\ll 1$, and seek an inner solution, $\F(Z)$, to equation (\ref{feq}). We approximate the coefficients by their Taylor series expansions about $\z=\zcd$, so that equation (\ref{feq}) becomes
\be
\frac{\short}{\delta^2 }\ \F''(Z)-\frac{2}{\delta}\ \F'(Z)  -\frac{\short}{\delta} \frac{\Uppc}{\Upc Z }\ \F(Z)=0\ ;\qquad Z=O(1), \label{balance}
\ee
where the subscript `c' denotes evaluation at the critical level, $\z=\zcd$. By balancing the two first terms, we obtain the distinguished limit $\delta = \short $ \citep{B-O}, and hence must solve
\be
\F''(Z) - 2 \F'(Z)=\short\ \frac{\Uppc}{\Upc Z }\ \F(Z),\qquad \short\to 0^+.
\ee
Because the boundary layer has an $O(\short)$ thickness, the appropriate inner expansion is 
\be
\F(Z)= F_0(Z) +\short\ F_1(Z) + O\big(\short^2\big), \qquad \short\to 0^+.
\ee
The general solutions are
\begin{align}
F_0(Z) &= A\ e^{2Z} + B,\qquad A,B \in \mathbb{C},\label{F_0}\\
\text{and}\qquad F_1(Z) &= \frac{\U''_{\rm{c}}}{\U'_{\rm{c}}}\ \int^Z_b dx\ e^{2x}  \int^x_a dt\ \frac{e^{-2t}}{t}\ F_0(t).\label{F_1}
\end{align}
Next we determine the integration constants, $A$ and $B$, and the bounds of integration, $a$ and $b$, by asymptotic matching. 

\subsection{Asymptotic matching and uniformly valid composite solutions}

We must match the inner solution and the two outer solutions, as $\short\to 0^+$. We use the superscripts `$\ell$' and `u' for lower and upper, respectively. 

The lower outer solution, $\fl(\z)$, is defined for $0~\le~\z~\ll~\zcd$, and must satisfy the lower boundary condition; 
\refstepcounter{equation}
$$
f_0^{\ell}(0)=1,\qquad\text{and}\qquad \fln(0)=0. \label{BCorder}
  \eqno{(\theequation{\mathit{a},\mathit{b}})}
$$
The upper outer solution, $\fu(\z)$, is defined for $\z\gg\zcd$ and hence must satisfy condition (\ref{newBCinf}\textit{a}). We use the following matching conditions, 
\begin{align}
\lim\limits_{\z\to \zcd^- }\fl(\z)&=\lim\limits_{Z\to-\infty}\F(Z),  \label{matchcond1}\\
 \text{and}\quad \lim\limits_{\z\to \zcd^+}\fu(z)&=\lim\limits_{Z\to+\infty}\F(Z), \label{matchcond2}
\end{align}
and then apply the Van Dyke additive rule \citep{B-O} to construct uniformly valid composite solutions;
\be
\text{uniform approx} = \text{inner}+\text{outer} - \text{common part}.  \label{vandyke}
\ee
We stress that this is possible because the lower and upper outer solutions happen to have a common analytical expression. 
\subsubsection{Leading order}
To leading order, the outer solutions are constant (see Eq. \ref{outershort}\textit{a}), and because of the boundary condition (\ref{BCorder}\textit{a}) we have
\be
f_0^{\ell}=1. \label{l0}
\ee
The leading order inner solution is given by equation (\ref{F_0}). To preempt divergence as $Z\to+\infty$, we impose $A=0$, and the matching condition (\ref{matchcond1}) implies that $B=1$. Therefore, 
\be
F_0 =1, \label{in0}
\ee
which, upon imposition of the matching condition (\ref{matchcond2}), yields
\be
f_0^{\rm{u}}=1.\label{u0}
\ee
We combine the solutions (\ref{l0}), (\ref{in0}) and (\ref{u0}) using the additive rule (\ref{vandyke}), and thus obtain a uniformly valid composite solution at leading order;
\be
\funif= 1+O(\short). \label{leadingunif}
\ee
The effect of the boundary layer appears only at the next order. 
\subsubsection{Order $\short $}    \label{matchingshort}
Following equation (\ref{outershort}\textit{b}), the lower and upper outer solutions at order $\short$ are 
\begin{align}
\fln(\z) &= -\frac{1}{2} \int^{\z}_{\zl} d\tilde{z}\   \frac{\U''(\tilde{z})} {  \U(\tilde{z})-\C_r }\ ,\\
\text{and}\quad  \fun(\z)&= -\frac{1}{2} \int^{\z}_{\zu} d\tilde{z}\   \frac{\U''(\tilde{z})} {  \U(\tilde{z})-\C_r }\ ,
\end{align}
respectively. From the Laurent series expansion
\be
\frac{\U''(\z)} {  \U(\z)-\C_r }= \frac{\Uppc}{\Upc (\z-\zcd)}+ \frac{1}{2}\bigg[\frac{\Uppc}{\Upc}\bigg]^2 + O(\z -\zcd),
\ee
we deduce that 
\begin{align}
\fln(\z) & \sim - \frac{\Uppc}{2\Upc} \Log\bigg(\frac{\z-\zcd  }{\zl-\zcd}\bigg),\quad \z\to\zcd^-, \label{innerl} \\
\text{and}\quad  \fun(\z) & \sim -\frac{\Uppc}{2\Upc} \Log\bigg(\frac{\z-\zcd  }{\zu-\zcd}\bigg),\quad \z\to\zcd^+, \label{inneru} 
\end{align}
where $\Log$ denotes a continuation of the natural logarithm to the negative real numbers, 
\be
\Log(\z-\zcd)=
\bc
\ln|\z-\zcd|- i\pi\quad\text{if }\Upc>0,\\
\ln|\z-\zcd|+ i\pi\quad\text{if }\Upc<0,
\ec
\quad \text{for }\z<\zcd. 
\label{Log}
\ee
The choice of the branch cut -- just above the negative real axis if $\Upc>0$, just below otherwise -- follows from \citet{linbook}.

From equation (\ref{F_1}), the inner solution at order $\short$ is
\be
F_1(Z) = \frac{\U''_{\rm{c}}}{\U'_{\rm{c}}}\ \int^Z_b dx\ e^{2x}  \int^x_a dt\ \frac{e^{-2t}}{t}.
\ee
We split the integral over $t$ as 
\be
 \int^x_a =  \int^x_{+\infty} + \int_a^{+\infty}.
\ee
Noting that the second integral is a constant, $C\in\mathbb{C}$, we obtain
 \be
F_1(Z)= -\frac{\Uppc}{\Upc}\bigg\{ \int^Z_b dx\ e^{2x}  E_1(2x)+ \frac{C}{2}\Big( e^{2Z}- e^{2b}\Big)\bigg\}. \label{F1}
\ee
where 
\be
E_1(x) \equiv\int_x^{+\infty} dt\ \frac{e^{-t}}{t}, \qquad |\arg(x)|<\pi, 
\ee
is the exponential integral.
For large values of $|x|$, the divergent series \citep{B-O}
\be
E_1(x)=\frac{e^{-x}}{x}\ \sum_{n=0}^{N-1} \frac{n!}{(-x)^n}\ ,\quad x\to \infty\ ;\quad |\arg(x)|<\frac{3\pi}{2},
\ee
yields
\be
e^{2x}  E_1(2x)\sim\frac{1}{2x}\ , \qquad x\to \pm\infty.
\ee
After integration, we find
\be
F_1(Z) \sim -\frac{\Uppc}{2\Upc}\bigg\{ \Log\bigg[  \frac{Z}{b} \bigg]+ C\Big( e^{2Z}- e^{2b}\Big)\bigg\}, \quad Z\to \pm\infty.\label{outerlim}
\ee
Recalling that $Z=(\z-\zcd)/\short$, the matching of the inner limits, given by equations (\ref{innerl}) and (\ref{inneru}), and the outer limits, given by equation (\ref{outerlim}), yields
\refstepcounter{equation}
$$
C = 0 \qquad\text{and}\qquad \zl-\zcd =\short b =  \zu-\zcd. \label{const}
  \eqno{(\theequation{\mathit{a},\mathit{b}})}
$$
Because the boundary condition (\ref{BCorder}\textit{b}) requires that 
\be
\zl=0, \label{zl}
\ee
the outer solution at order $\short$ has the same expression in the lower and upper outer regions;
\be
f_1(\z) = -\frac{1}{2} \int_0^{\z} d\tilde{z}\   \frac{\U''(\tilde{z})} {  \U(\tilde{z})-\C_r }\ , \quad \z\ll\zcd\ \text{ and }\ \z\gg\zcd. \label{outer}
\ee
The lower outer solution is real, because the path of integration along the real axis does not reach the branch point $\z=\zcd$. However, for the upper outer solution we have to make a detour in the complex plane. We deform the path of integration in the upper part if $\Upc>0$, and in the lower part otherwise. In Appendix \ref{globprop}, we show that
\be
\Im\{\fun\} = -\frac{\pi}{2}\ \frac{\Uppc}{|\Upc|}\ . \label{imupper}
\ee
Using the matching conditions (\ref{const}) together with equation (\ref{zl}), the inner solution~(\ref{F1}) becomes 
\be
F_1(Z)= -\frac{\Uppc}{\Upc} \int^Z_{-\zcd/\short} dx\ e^{2x}  E_1(2x). \label{inner}
\ee
Finally, we use the Van Dyke rule (\ref{vandyke}) to construct a uniformly valid composite solution at order $\short$; 
\begin{widetext}
\be
\funiff(\z) = 1+ \short\Bigg\{-\frac{\Uppc}{\Upc}\ \int^Z_{-\zcd/\short} dx\ e^{2x}  E_1(2x)-\frac{1}{2} \int_0^\z  d\tilde{z} \  \frac{\U''(\tilde{z})} {  \U(\tilde{z})-\C_r }+ \frac{\Uppc}{2\Upc}\ \Log\bigg( 1- \frac{\z}{\zcd} \bigg)\Bigg\}+O\big(\short^2\big). \label{nextunif}
\ee
\end{widetext}
In Appendix \ref{globprop}, we verify that $\chi(\z) = e^{-\kd\z}\funiff(\z)$ satisfies  the global property; 
\be
\Im\big\{\chi'(0^+)\big\} =-\pi\ \frac{\Uppc}{|\Upc|}\ |\crit|^2 ,  \label{property}
\ee
at leading order.
In Appendix \ref{solcrit} we show that 
\be
f(\zcd)  = 1 +\frac{\short}{2}\ \frac{\Uppc}{\Upc} \Bigg(\euler +\ln\bigg[\frac{2\zcd}{\short}\bigg] - i \pi \Bigg) +O\big(\short^2\big). \label{fcrit}
\ee
Equation (\ref{fcrit}) generalizes a result that Miles derived in an Appendix to \citet{morland-saffman} using an exact solution of the Rayleigh equation for an exponential wind profile. 
\subsection{Comparison with the numerical solution}
\begin{figure*}[htbp!]
        (a)\includegraphics[trim = 0 0 0 0, clip, width = 0.46\textwidth]{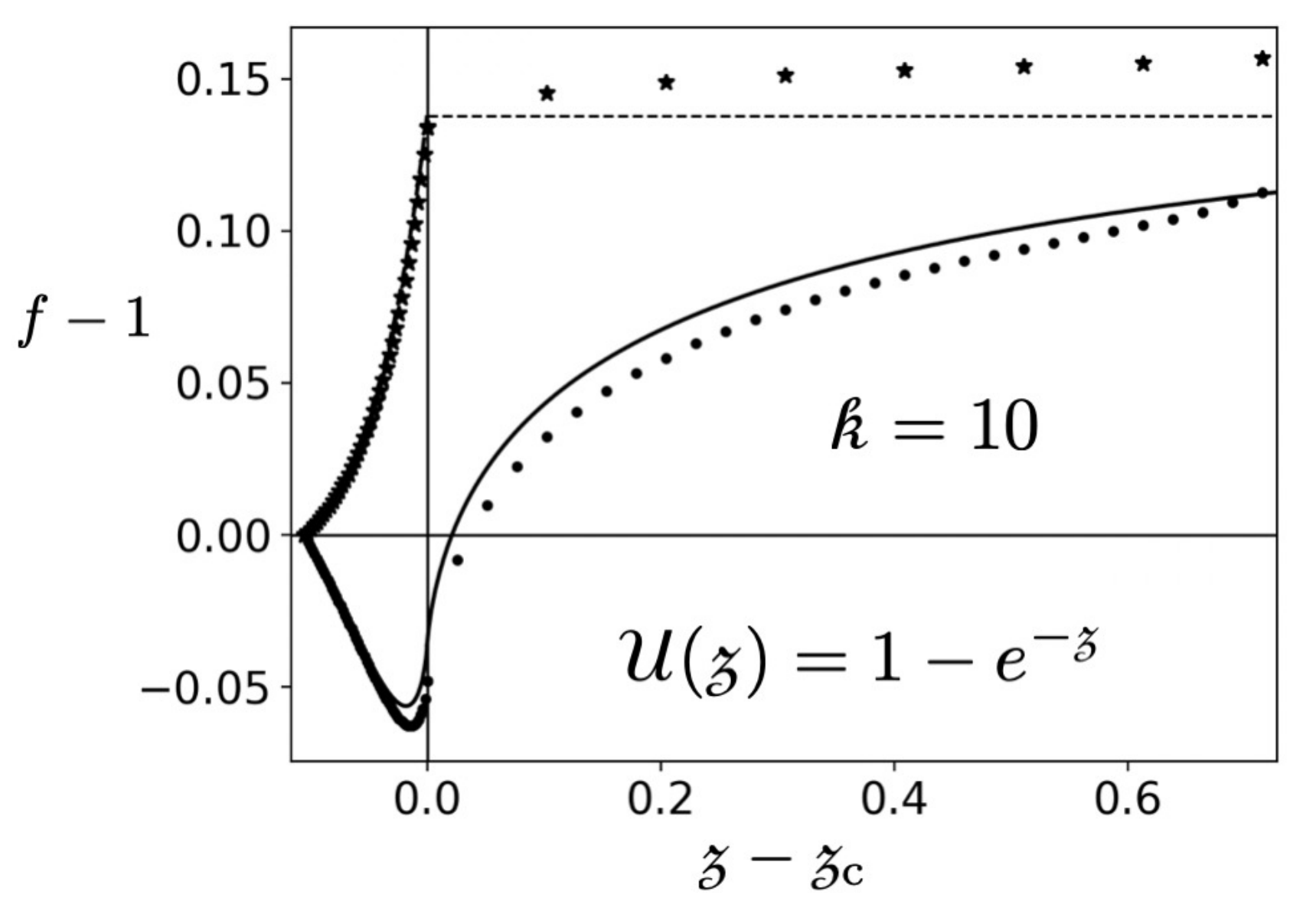}	  	
        (b)\includegraphics[trim = 0 0 0 0, clip, width = 0.47\textwidth]{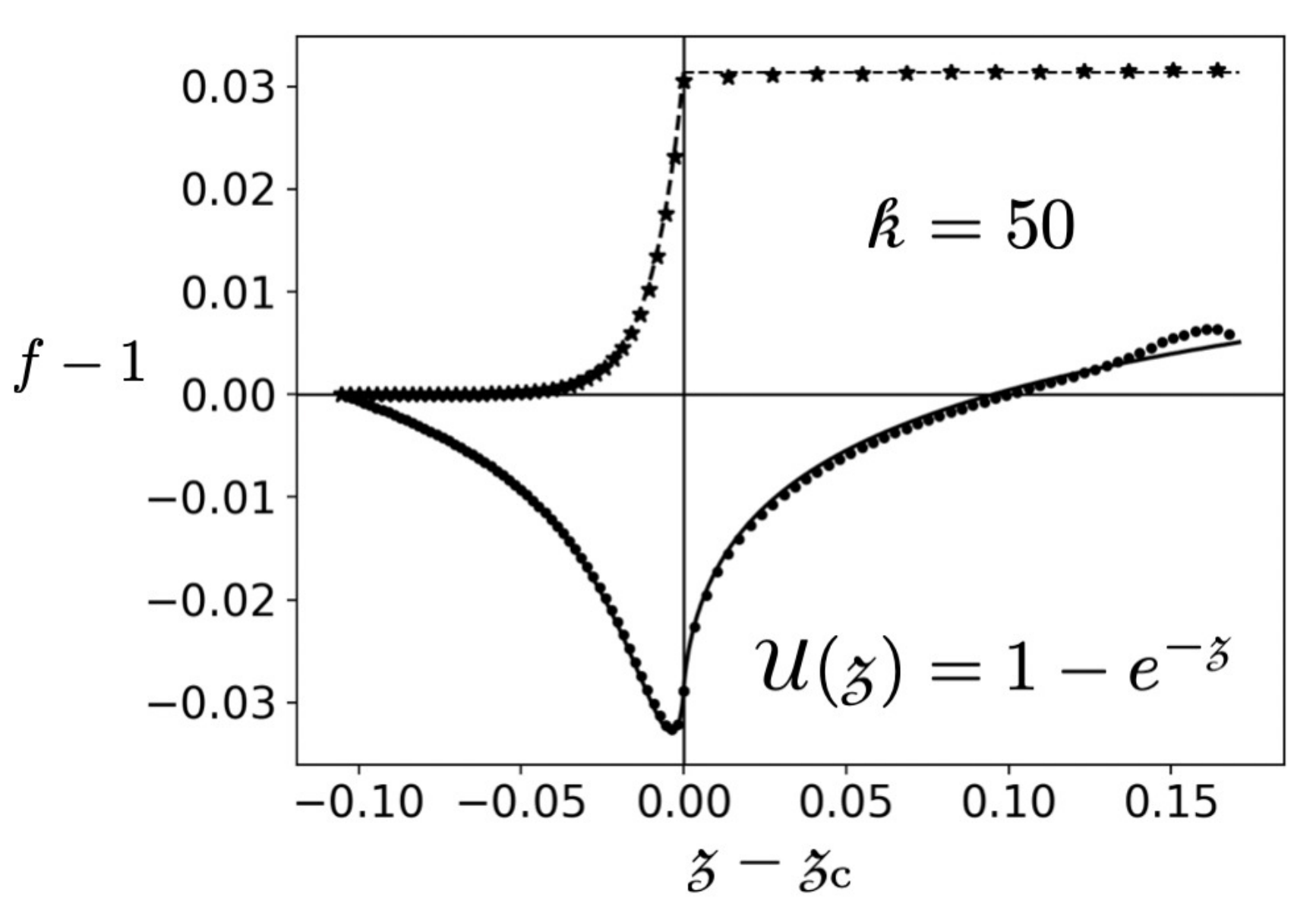}
        (c)\includegraphics[trim = 0 0 0 0, clip, width = 0.45\textwidth]{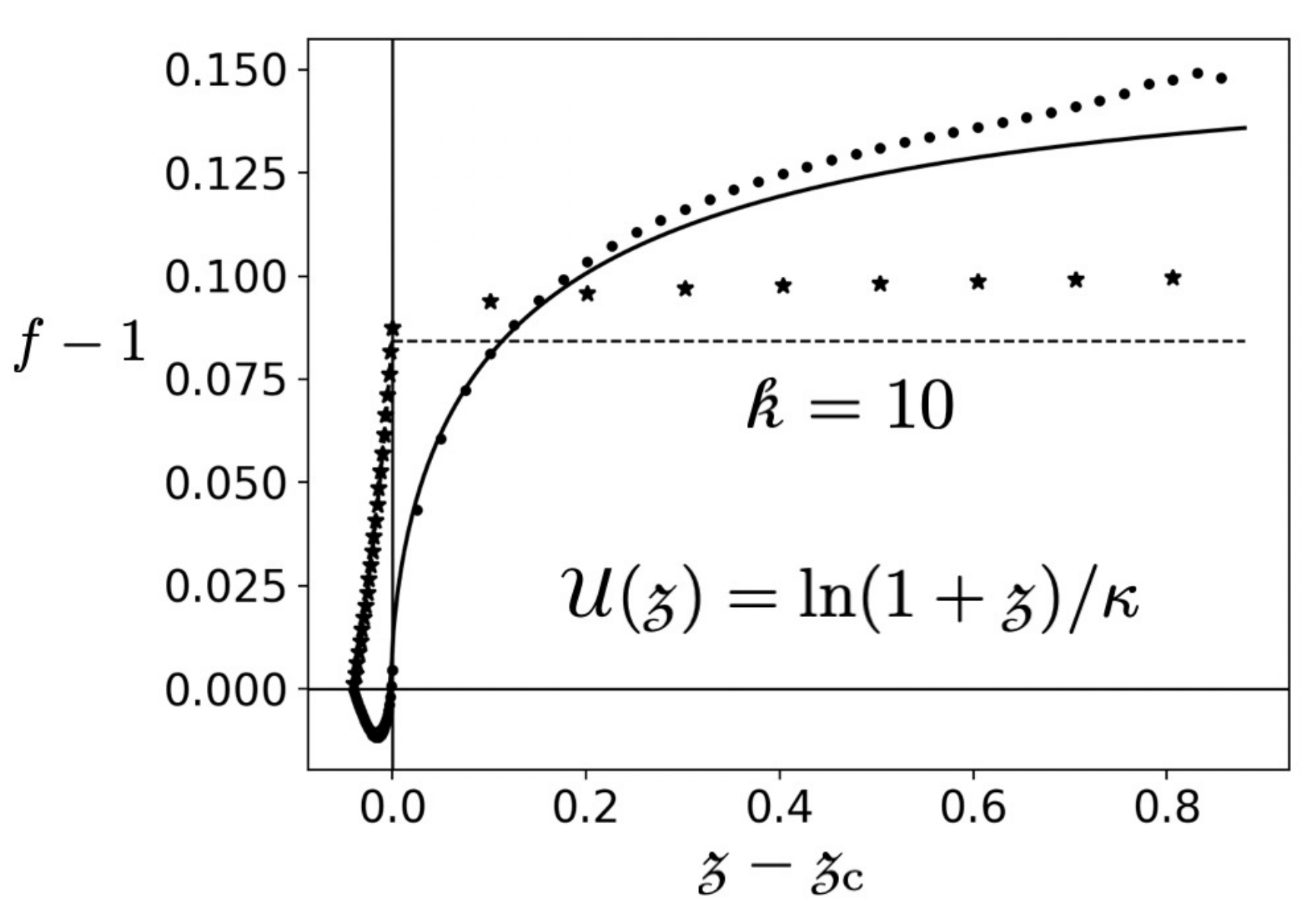}	  	
        (d)\includegraphics[trim = 0 0 0 0, clip, width = 0.49\textwidth]{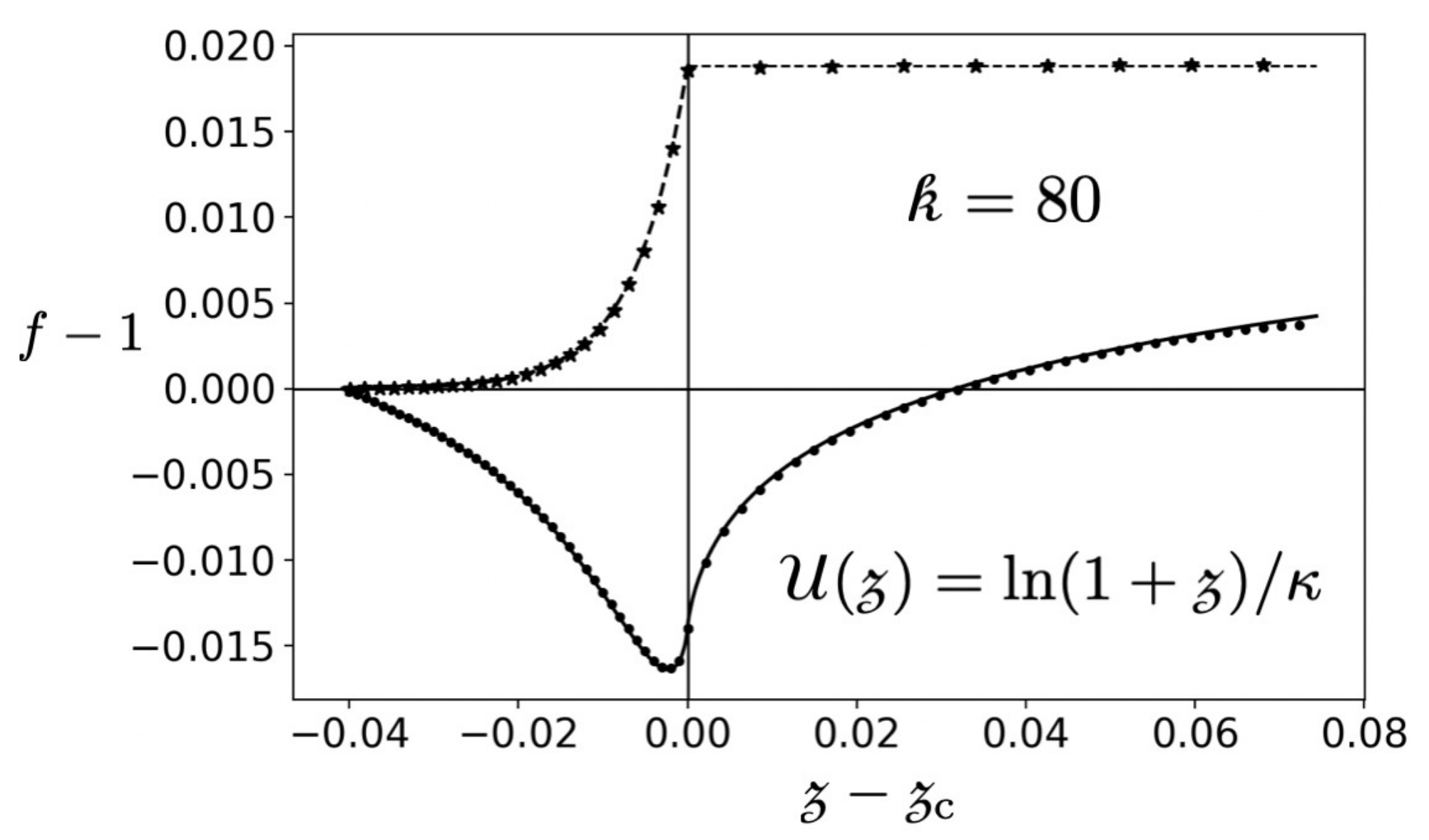}
  \caption{Comparison of the uniformly valid composite solution (\ref{nextunif}) with the numerical solution of the Rayleigh equation for the exponential (a,b) and logarithmic (c,d) wind profiles, with $\C_r=0.1$ and different values of $\kd$. The dots and the stars denote the real and imaginary parts of the numerical solution, respectively. The solid line shows the real part of (\ref{nextunif}) and the dashed line the imaginary part.}
\label{BLtheory}
\end{figure*}
We numerically solve equation (\ref{feq}) for the two standard wind profiles
\refstepcounter{equation}
$$
\U(\z) = 1-e^{-\z} \qquad\text{and}\qquad  \U(\z) = \ln(1+\z)/\kappa,
  \eqno{(\theequation{\mathit{a},\mathit{b}})}
$$
where, as before, $\kappa=0.4$ is the von K\'arm\'an constant. 
We compare the numerical solution with our uniformly valid composite solution (\ref{nextunif}) for fixed values of $\kd$ and $\C_r$ in Figure  \ref{BLtheory}.  (Note that any dispersion relation can be retrieved with a proper choice of the control parameters.)  For both profiles the composite solution and the numerical solution agree very well. We distinguish the lower (upper) outer solutions with their imaginary part being equal to zero (a positive constant). Consistent with the Frobenius solution \citep{drazin-reid}, the solution within the inner layer depends on the wind profile and the dispersion relation solely through the scale factor $\Uppc/\Upc$ and the bound of integration $\zcd/\short$. Since the phase of the solution of the Rayleigh equation changes only within the boundary layer, we conclude that the interaction of short waves with the wind principally occurs therein. 
In contrast, we showed that for $\kd\ll1$ the phase varies from $\z=0$ to $\z=\zcd$, so that long waves interact with the wind all the way from the mean water surface to the critical level~\citep{bonfils-et-al22}. 

\subsection{Similarity solution}

We have non-dimensionalized the variables using external parameters characterizing the shear in the air. However, the general short wave solution (\ref{nextunif}) can also be written in terms of $\xi\equiv k z$ and $\Usim\equiv U/\tilde{c}_r$, where $\tilde{c}_r$ is the dimensional version of $\C_r$. The wind-generated ripples have a continuous wavenumber spectrum so that $k$, and subsequently $\tilde{c}_r$, are variables rather than parameters. In that sense, $\xi$ is a similarity variable introduced by \citet{miles57}, and the Rayleigh equation has a self-similar solution, $\phi=\phi(\xi)$, which for short waves is
\begin{widetext}
\be
\phi(\xi) = e^{-\xi} \Bigg\{  1-\frac{\Usim''(\xic)}{\Usim'(\xic)}\ \int^{\xi-\xic}_{-\xic} dx\ e^{2x}  E_1(2x)-\frac{1}{2} \int_0^\xi  d\tilde{z} \  \frac{\Usim''(\tilde{z})} {  \Usim(\tilde{z})-1 }+ \frac{\Usim''(\xic)}{2\Usim'(\xic)}\ \Log\bigg( 1- \frac{\xi}{\xic} \bigg)\Bigg\}.
\ee
\end{widetext}
\begin{figure}
  \centerline{\includegraphics[scale=0.25]{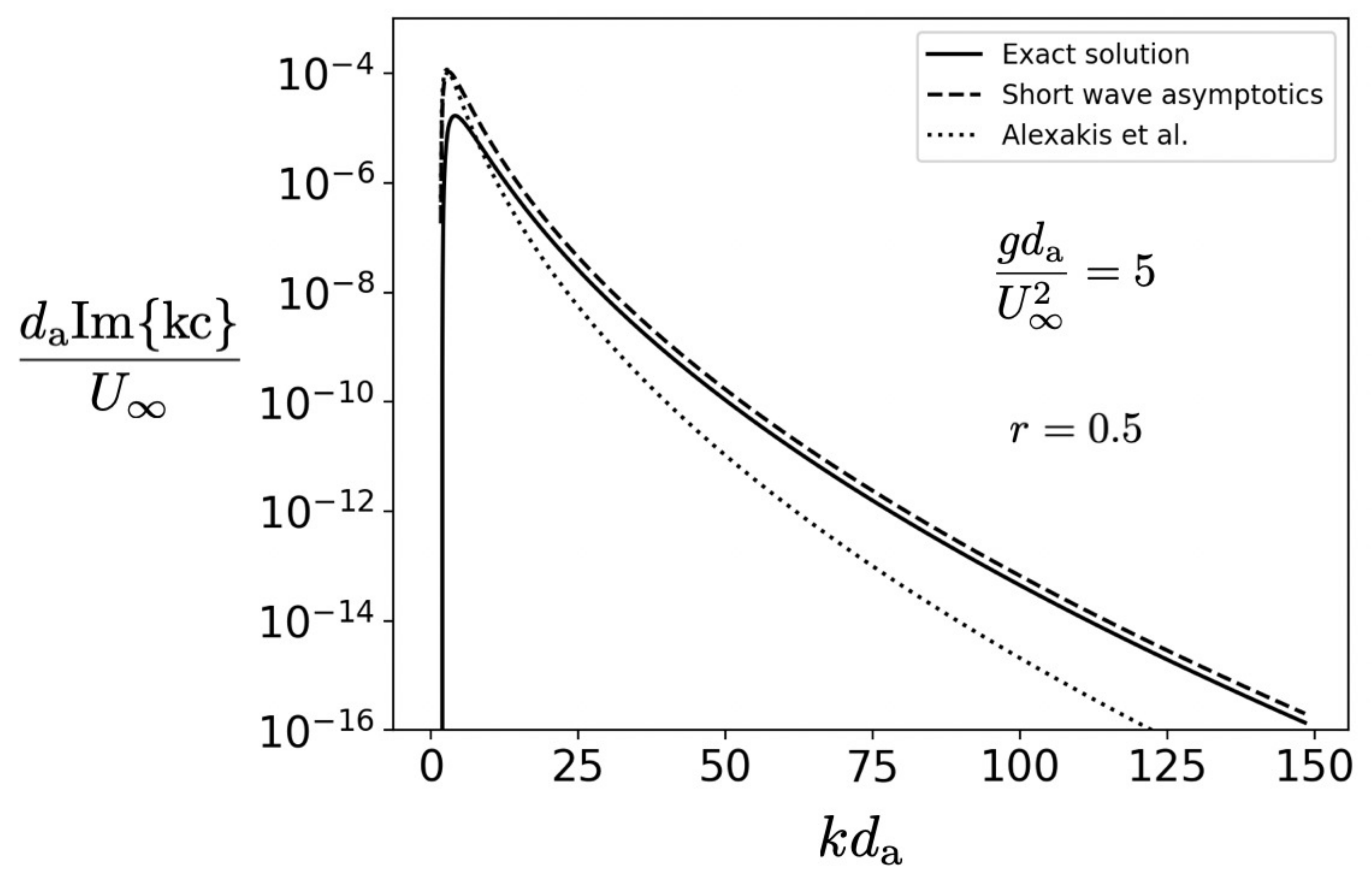}}
  \caption{Comparison of the growth rate of gravity waves obtained by \citet{alexakis-et-al04} with the predictions of our short wavelength asymptotic treatment (cf. Eq. \ref{gr+}) and the growth rate calculated from the exact (hypergeometric) solution for a wind profile $U(z) = U_\infty\ (1-e^{-z/\da})$ and a density ratio $r=0.5$.}
\label{alexcomparison}
\end{figure}
\section{The WKB method in the shortwavelength limit} \label{comment}
\citet{alexakis-et-al04} proposed a solution of the Rayleigh equation for short waves using the WKB method. However, because their solution does not satisfy the global property~(\ref{property}), their growth rate has an extra factor $\pi/\kd$; see their equation (A31). In Figure \ref{alexcomparison}, we show the growth rate of gravity waves for an exponential wind profile and compare the results of \citet{alexakis-et-al04} with our short wave asymptotic solution and the exact solution.  Clearly the WKB method deviates from the others for all $k d_a \gtrsim 25$, and in what follows we explain the origin of the deviation.

\begin{figure}
  \centerline{\includegraphics[scale=0.4]{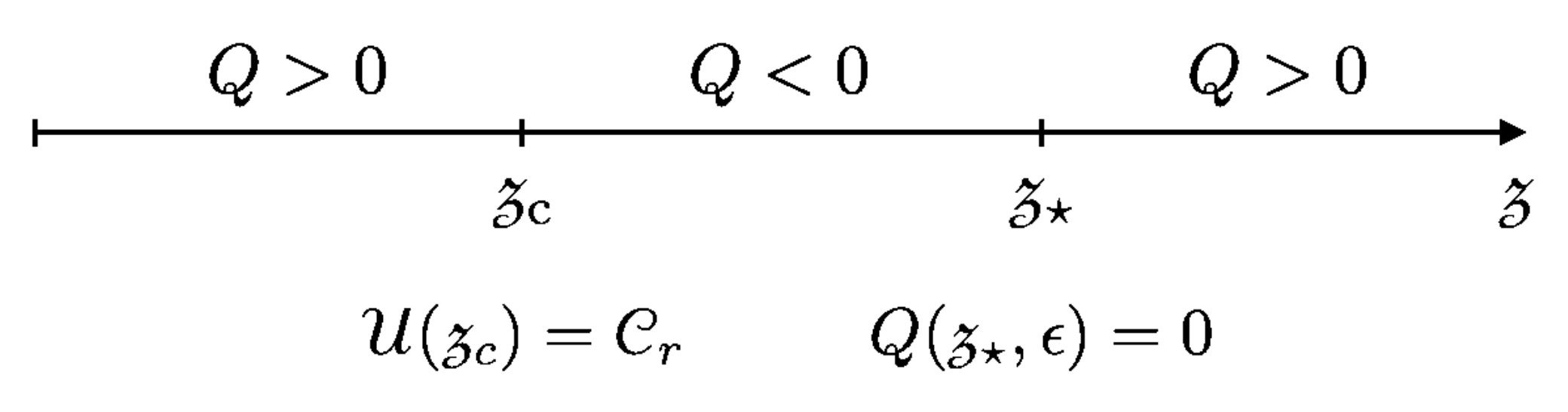}}
  \caption{The three intervals in the WKB approach of \citet{alexakis-et-al04} where $\zt$ is a simple turning point. }
\label{intervals}
\end{figure}

Setting $\short=1/\kd$, we write the Rayleigh equation in the form 
\be
\short^2 \chi''(\z)= Q(\z,\short) \chi(\z),\quad\text{with}\quad Q(\z,\short)\equiv 1+ \short^2\ \frac{\U''(\z)}{\U(\z)-\C_r}\ . \label{Q}
\ee
A WKB expansion is \citep{B-O}
\be
\chi(\z)\sim \exp\bigg[\frac{1}{\short}\ \sum_{n=0}^{+\infty} \short^n S_n(\z)\bigg], \qquad \short\to0. \label{WKB}
\ee
In the standard case, $Q(\z,\short)$ does not depend on the small parameter $\short$. Then, we can take only two terms in the series  (\ref{WKB}): this, physical optics approximation, yields
\begin{widetext}
\be
\chi(\z)\sim C_1 \big[Q(\z)\big]^{-\frac{1}{4}}\exp\bigg[\frac{1}{\short}\int_a^\z d\tilde{z} \sqrt{Q(\tilde{z})}\bigg]+ C_2 \big[Q(\z)\big]^{-\frac{1}{4}}\exp\bigg[-\frac{1}{\short}\int_a^\z d\tilde{z} \sqrt{Q(\tilde{z})}\bigg] \label{WKBpos}
\ee
for $Q(\z)>0$, and 
\be
\chi(\z)\sim C_3 \big[-Q(\z)\big]^{-\frac{1}{4}}\exp\bigg[\frac{i}{\short}\int_b^\z d\tilde{z} \sqrt{-Q(\tilde{z})}\bigg]+ C_4 \big[-Q(\z)\big]^{-\frac{1}{4}}\exp\bigg[-\frac{i}{\short}\int_b^\z d\tilde{z} \sqrt{-Q(\tilde{z})}\bigg],
\label{WKBneg}
\ee
for $Q(\z)<0$, where the bounds of integration, $a$ and $b$, are arbitrary.
\end{widetext}
For $\U''(\z)<0$ and $Q(\z,\short)$ introduced in equation (\ref{Q}), \citet{alexakis-et-al04} used the physical optics approximation on three intervals, defined in Figure \ref{intervals}, and obtained three solutions of the form of (\ref{WKBpos}) or (\ref{WKBneg}), in which they neglected the fact that $\short\ll1$ within $Q(\z,\short)$. Then, they had to match those solutions in order to determine the integration constants $C_1$, $C_2$, etc. The common matching procedure takes place at the simple turning point $\zt$ \citep[e.g.][]{B-O}. However, they needed inner solutions near the critical level, $\zcd$, and it is at this juncture that the appeal of the WKB method is understood. Indeed, they found that the solution of the Rayleigh equation near the singularity can be represented in terms of Bessel functions for $\z>\zcd$, and in terms of modified Bessel functions for  $\z<\zcd$. Moreover, the outer limits of those inner solutions formally match the inner limits of the physical optics approximations. Unfortunately, however, the distance between the critical point $\zcd$ and the turning point $\zt$ actually shrinks as $\short$ tends to zero. For instance, in the case of the exponential profile $\U(\z)=1-e^{-\z}$, we have
\be
\zt-\zcd = \ln(1+\short). 
\ee
Hence, the interval where $Q<0$ cannot be taken as an outer region. In fact, as seen in Figure \ref{BLtheory}, the numerical solution of the Rayleigh equation does not exhibit oscillatory behavior. 

In the vicinity of the critical level, $\zcd$, we introduce the inner variable
\be
\Z=-\frac{\Uppc}{\Upc}\ (\z-\zcd),
\ee
in terms of which the inner solutions are 
\begin{align}
\chi_{\text{in+}}(\Z>0) &= \sqrt{\Z}\Big[ A\ J_1\big(2\sqrt{\Z}\big)+ B\ Y_1\big(2\sqrt{\Z}\big)\Big],\\
\chi_{\text{in-}}(\Z<0) &= \sqrt{-\Z}\Big[ C\ I_1\big(2\sqrt{-\Z}\big)+ D\ K_1\big(2\sqrt{-\Z}\big)\Big],
\end{align}
where $A$, $B$, $C$, and $D$ are complex constants, and $J_1$ and $Y_1$ ( $I_1$ and  $K_1$) are the Bessel functions (modified Bessel functions) of order $1$. For the solution to be continuous and its derivative to have the correct behavior at $\z=\zcd$, we must establish some relationships between these constants. Although this was not done explicitly by \citet{alexakis-et-al04}, we could use their final matching formulae to determine such relations to find them to be incompatible with the following; 
\be
I_1(\z)= -i\ J_1(i \z) \ \text{ and }\ K_1(\z)= -\frac{\pi}{2}\Big[J_1(-i\z)-i\ Y_1(-i\z)\Big],
\ee
for $-\pi<\arg(\z)\le \pi/2$. We find that
\be 
C = -A+i\ B\qquad\text{and}\qquad D = -\frac{2}{\pi}\ B. \label{relations}
\ee
To check the consistency of these results, we use equation (\ref{relations}) for the numerical integration of the Rayleigh equation and retrieve the growth rates calculated by \citet{beji-nadaoka}. 
\section{Conclusion}\label{ccl}

We have studied the effect of a shear flow on interfacial capillary-gravity waves when their wavelength is much smaller than the characteristic length scale of the flows in the fluids bounding the interface. Using the dimensionless inverse wavenumber as a small parameter, $\short = 1/\kd$, we asymptotically solved the eigenvalue problem for the stability of an arbitrary parallel flow $\bb{U}=U(z)\ \bb{\hat{x}}$ through a two-fluid interface with a density ratio that is not necessarily small. We constructed uniformly valid composite solutions of the governing Rayleigh equation, where the real part of the eigenvalue is a power series in $\short$. We showed that including the effect of surface tension changes the nature of the leading order solution, and that exponentially small terms must be considered in order to have a non-zero imaginary part of the solution. From a physical view-point, there is a prograde mode whose phase speed is greater than the speed of the water surface, $\Us$, and a retrograde mode whose phase speed is smaller than $\Us$. When the velocity of the shear flow equals the phase speed of one these modes, there is a critical layer where the flow transfers energy to the wave. If the critical layer is in the air (water), this is called the Miles (rippling) instability. The only case considered in the literature thus far is when the prograde mode undergoes the Miles instability and the retrograde mode undergoes the rippling instability. In \S~\ref{doublecrit} we studied the situation where the prograde mode can undergo both instabilities and the retrograde mode is neutral. In the short wave limit, we found that (i) the effect of the shear on the phase speed of the two modes depends only on the derivatives of $U$ at $z=0^\pm$; and (ii) the Miles and rippling instabilities have a growth rate of the same form.  Indeed, the interaction between the shear flow and the waves is mostly reduced to a narrow region around the critical level, an internal boundary layer of thickness $\short$ where the solution of the Rayleigh equation has a self-similar structure. Heuristically speaking, the waves are barely influenced by the flow outside of this region. Nonetheless, we showed that there are signifiant effects on dispersion and growth rates when the characteristic velocity of the shear flow is large and the density ratio is close to $1$. Finally, we showed how the WKB approach of solving the Rayleigh equation breaks down.  
\section*{Acknowledgements}
We thank Cristobal Arratia for suggesting the possibility of a prograde mode with two critical layers, developed in \S \ref{doublecrit}. A. F. Bonfils thanks Neil Balmforth and Robert Rosner for discussions of the WKB approach and 
the 2022 Geophysical Fluid Dynamics Summer Study Program at the Woods Hole Oceanographic Institution, which is supported by the National Science Foundation and the Office of Naval Research under No. OCE-1332750. All authors acknowledge Swedish Research Council grant no. 638-2013-9243. Nordita is supported in part by NordForsk. 
\appendix

\section{Solution of equation (\ref{feq}) at the singular point $\z=\zcd$}\label{solcrit}
The regular singular point $\z=\zcd$ is located in the inner region where the solution at order $\short$ is given by equation (\ref{inner}), which we reproduce here;
\be
F_1(Z)= -\frac{\Uppc}{\Upc} \int^Z_{-\zcd/\short} dx\ e^{2x}  E_1(2x),\ \text{ where }\ Z=\frac{\z-\zcd}{\short}\label{inner-rep}
\ee
is the inner variable. 
\subsection{Imaginary part}
From the series representation of the exponential integral 
\be
E_1(x) = -\euler-\Log(x)-\sum_{n=1}^{+\infty} \frac{(-x)^n}{n\ n!}, \quad |\arg(x)|<\pi, \label{E1}
\ee
where $\euler=0.577$ is the Euler constant, we see that $x=0$ is a logarithmic branch point for the integral in equation (\ref{inner-rep}). The only contribution to the imaginary part of $F_1$ arises from the integration of $\Log (x)$ along the branch cut. Namely, if $\step$ is the Heaviside step function then 
\begin{widetext}
\be
\Im\bigg\{\int^Z_{-\zcd/\short} dx\ e^{2x}  E_1(2x)\bigg\} = \pm\step(-Z)\ \pi\int^Z_{-\zcd/\short} dx\ e^{2x} \pm \step(Z) \ \pi \int^0_{-\zcd/\short} dx\ e^{2x} , \label{imag}
\ee
\end{widetext}
where we select the plus sign when $\Upc>0$, and the branch cut is just above the negative real axis, and the minus sign otherwise. From equation (\ref{imag}), we find
\be
\Im\big\{F_1(Z)\big\} =
\bc
  -\frac{\pi}{2}\ \frac{\Uppc}{|\Upc|}   \big(e^{2Z}-e^{-2\zcd/\short}\big)\quad\text{if } Z<0,\\
  \\
   -\frac{\pi}{2}\ \frac{\Uppc}{|\Upc|}  \big(1-e^{-2\zcd/\short}\big)\quad\text{if } Z\ge 0.
 \ec \label{F1im}
\ee
Then, upon taking the limit $Z\to0$ we obtain
\be
\Im\big\{f(\zcd)\big\} = -\short\ \frac{\pi}{2}\ \frac{\Uppc}{|\Upc|}  \Big(1-e^{-2\zcd/\short}\Big) + O\big(\short^2\big).
\ee
It is interesting to take the outer limits $Z\to\pm\infty$ of the expression (\ref{F1im}). In particular, if the matching was done properly, these should be equal to the inner limits of the imaginary parts of the lower and upper outer solutions, viz.,
\begin{align}
\lim\limits_{\z\to\zcd^-}\Im\big\{ \fln(\z)\big\} &= \frac{\pi}{2}\ \frac{\Uppc}{|\Upc|}\ e^{-2\zcd/\short}, \\
\text{and}\quad \lim\limits_{\z\to\zcd^+}\Im\big\{ \fun(\z)\big\} &=  -\frac{\pi}{2}\ \frac{\Uppc}{|\Upc|}  \Big(1-e^{-2\zcd/\short}\Big). 
\end{align}
In each case, we shall check that
\be
\lim\limits_{\short\to0^+} \frac{\zcd(\short)}{\short} = +\infty. \label{zc/epsilon}, 
\ee
which is the condition of separation of the boundary layer of thickness $\short$ from the lower boundary. Thus, we consistently retrieve a zero imaginary part in the lower outer region. We recall that the upper outer solution is given by (Eq. \ref{outer}) 
\be
\fun(\z) = -\frac{1}{2} \int_0^{\z} d\tilde{z}\   \frac{\U''(\tilde{z})} {  \U(\tilde{z})-\C_r }\ , \qquad \z\gg\zcd. 
\ee
According to the Sokhotski-Plemelj theorem, 
\begin{widetext}
\be
\lim\limits_{\C_i\to0^+} -\frac{1}{2} \int_0^{\z} d\tilde{z}\   \frac{\U''(\tilde{z})} {  \U(\tilde{z})-(\C_r+i \C_i) } = -\frac{1}{2}\ \cauchy  \int_0^{\z} d\tilde{z}\   \frac{\U''(\tilde{z})} {  \U(\tilde{z})-\C_r } - i\pi\ \frac{\Uppc}{2|\Upc|}\ ,\qquad \forall\z>\zcd,
\ee
\end{widetext}
where $\cauchy$ denotes the Cauchy principal value. Hence, the imaginary part in the upper outer region is constant, equal to the upper outer limit of $\Im\{F_1\}$ up to exponentially small terms. 
\subsection{Real part}
\begin{figure}
  \centerline{\includegraphics[scale=0.22]{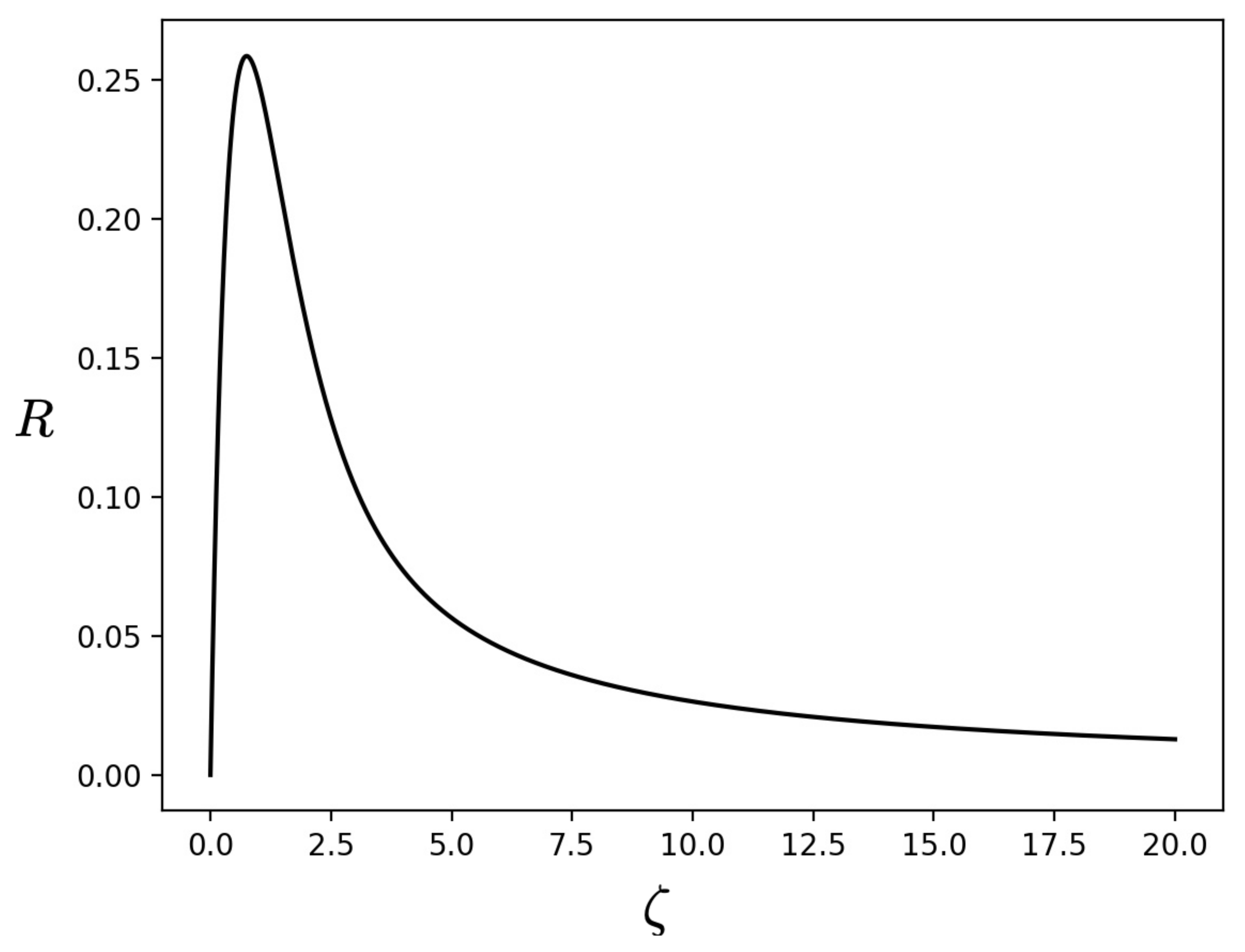}}
  \caption{The `Remainder' $R(\zeta)$ given by equation (\ref{R}).}
\label{remainder}
\end{figure}
At the leading order, we have
\be
f(\zcd) =1 +O(\short). \label{limcrit}
\ee
Here we calculate the next order contribution to the real part. For convenience, we define
\be
J(Z) \equiv\Re\bigg\{\int^Z_{-\zcd/\short} dx\ e^{2x}  E_1(2x)\bigg\}, 
\ee
and we express the exponential integral using the Ramanujan series as
\be
E_1(x) = -\euler-\Log(x)+ e^{-\frac{x}{2}}\sum_{n=1}^{+\infty} \frac{s_n}{n!} \bigg[\frac{x}{2}\bigg]^n,\quad  s_n =  \sum_{j=0}^{\left\lfloor \frac{n-1}{2} \right\rfloor} \frac{1}{j+\frac{1}{2}}\ , \label{ramanujan}
\ee
where $\left\lfloor x \right \rfloor$ denotes the floor function. This series converges faster than (\ref{E1}). We perform a direct integration as
\be
J(Z) = -\int^Z_{-\zcd/\short} dx\  e^{2x} \big(\euler + \ln(2) + \ln|x| \big) + \sum_{n=1}^{+\infty} \frac{s_n}{n!} \int^Z_{-\zcd/\short} dx\ e^x x^n,
\ee
and after some algebra we obtain 
\begin{widetext}
\be
J(Z) = \frac{1}{2}\Bigg\{ \Big[\euler+ \ln(2)\Big] \Big[e^{-2\zcd/\short} - e^{2Z}\Big]+ e^{-2\zcd/\short} \ln\bigg[\frac{\zcd}{\short}\bigg]-e^{2Z} \ln|Z|+E_1\bigg[\frac{2\zcd}{\short}\bigg]-E_1\big(2|Z|\big) \Bigg\}+ S(Z),
\ee
\end{widetext}
with
\be
S(Z) \equiv \sum_{n=1}^{+\infty} (-1)^n  \frac{\Gamma(n+1,-Z)- \Gamma(n+1,\zcd/\short)  }{ n!}\  s_n ,
\ee
in which
\be
\Gamma(a,x) \equiv \int_x^{+\infty} dt\ t^{a-1} e^{-t},
\ee
is the upper incomplete gamma function. Note that
\refstepcounter{equation}
$$
\Gamma(n+1,0) = n!\quad\text{and}\quad s_{2p-1}=s_{2p},\quad \forall p\in \mathbb{N}^{\star},
  \eqno{(\theequation{\mathit{a},\mathit{b}})}
$$
and that
\begin{widetext}
\be
\lim\limits_{Z\to 0} E_1(2\zcd/\short)-E_1\big(2|Z|\big) -e^{2Z} \ln|Z| = -\ln\bigg[\frac{\zcd}{\short}\bigg] + e^{-\zcd/\short}\sum_{n=1}^{+\infty} \bigg[\frac{\zcd}{\short}\bigg]^n \frac{ s_n}{n!}\ .
\ee
\end{widetext}
Therefore, 
\be
\lim\limits_{Z\to 0}  J(Z) = - \frac{1}{2} \Bigg(\euler + \ln\bigg[\frac{2\zcd}{\short}\bigg]\Bigg) \Big(1- e^{-2\zcd/\short} \Big) + R\bigg[\frac{\zcd}{\short}\bigg].
\ee
We define the `remainder' as
\be
R(\zeta) \equiv \frac{e^{-\zeta}}{2} \sum_{n=1}^{+\infty} \frac{\zeta^n}{n!}\ s_n - \sum_{n=1}^{+\infty} (-1)^n  \frac{\Gamma(n+1,\zeta)  }{ n!}\  s_n, \label{R}
\ee
which is a negligible quantity for $\zeta\gg 1$, as shown in Figure \ref{remainder}. Discarding the exponentially small terms, we arrive at
\be
\Re\big\{f(\zcd)\big\}  = 1 +\frac{\short}{2}\ \frac{\Uppc}{\Upc} \Bigg(\euler +\ln\bigg[\frac{2\zcd}{\short}\bigg]\Bigg) +O\big(\short^2\big).
\ee

\section{Global property of the solution of the Rayleigh equation}\label{globprop}

The dimensionless Rayleigh equation is
\be
\chi''(\z)+\bigg[\kd^2+\frac{\U''(\z)}{\U(\z)-\C_r}\bigg] \chi(\z)=0,  \label{rayl}
\ee
where $\C_r=\C_r(\kd)$ is a known function; the boundary conditions are 
\be
 \chi(0)=1\quad\text{and}\quad \lim\limits_{\z\to+\infty}\chi(\z)=0. \label{BCrayl}
\ee
Assuming that there is a unique level $\zcd>0$ such that $\U(\zcd)=\C_r$, \citet{miles57} showed that 
\be
\Im\big\{\chi'(0^+)\big\} =-\pi\ \frac{\Uppc}{|\Upc|}\ |\crit|^2, \label{miles}
\ee
where the subscript `c' denotes evaluation at $\z=\zcd$. The derivation of the global property (\ref{miles}) is analogous to the canonical derivation of Rayleigh's inflection point theorem as follows. Multiply the Rayleigh equation (\ref{rayl}) by the complex conjugate of $\chi(\z)$, integrate by parts, use the boundary conditions (\ref{BCrayl}) to evaluate the integrated term, and take the imaginary part; the result follows from the Sokhotski-Plemelj theorem. 

Here, we show that our asymptotic solution for short waves satisfies equation (\ref{miles}).  For $\z>0$, we let 
\be
\chi(\z) = e^{-\z/\short}\ f(\z),\qquad\text{with}\qquad \short=1/\kd\ . 
\ee
We take the derivative of the composite solution, $\funiff(\z)$, obtained in equation (\ref{nextunif})
and obtain 
\begin{widetext}
\be
\funiff'(0^+) = -\frac{\Uppc}{\Upc}  \bigg[E_1\bigg(\frac{2\zcd}{\short}\bigg)\pm i\pi\bigg] e^{-2\zcd/\short} +   \short\ \frac{\U''(0^+)}{2\C_{r}(\short)} -\frac{\Uppc}{\Upc}\ \frac{\short}{2\zcd}\ , \label{derunif}
\ee
\end{widetext}
where we select $+ i\pi$ if $\Upc>0$, and $- i\pi$ otherwise. The contribution to the real part of the exponential integral multiplied by the decaying exponential is negligible, and hence is discarded in equations (\ref{fpro}) and (\ref{fretro}). 

We use equation (\ref{derunif}) to calculate the left-hand side of the identity (\ref{miles});
\be
\Im\big\{\chi'(0^+)\big\}= \Im\big\{\funiff'(0^+)\big\} + \text{h.o.t.} = -\pi\ \frac{\Uppc}{|\Upc|}\ e^{-2\zcd/\short} . \label{left}
\ee
Using the results of Appendix \ref{solcrit}, we have
\be
\crit = e^{-\zcd/\short} \big\{1+O(\short)\big\}. \label{chicrit}
\ee
Using equation (\ref{chicrit}) in the right-hand side of equation (\ref{miles}), we recover the left-hand side of equation (\ref{left}).  

\section{Short wavelength cut-off of the rippling instability}\label{cutoff}

As shown in Figure \ref{rippling}, the rippling instability has a short wavelength or high wavenumber cut-off when the effect of surface tension is taken into account. Following \citet{young-wolfe}, we denote this cut-off $k^+_{\rm{neut}}$ where the subscript `neut' denotes neutral. In other words, the wave with wavenumber $k^+_{\rm{neut}}$ is marginally stable because its phase speed equals the lower bound of the velocity profile in the water, namely  
\be
c_-(k^+_{\rm{neut}}) = \lim\limits_{z\to -\infty} U(z)\equiv U_{-\infty},
\ee
where all variables are dimensional. Hence, using equation (\ref{cap-grav}) for the phase speed of short waves, $k^+_{\rm{neut}}~L~\equiv~\kd^+_{\rm{neut}}$ is solution of 
\be
\frac{\Us}{V} - \sqrt{\frac{\s\kd}{1+r}} + \frac{r\U'(0^+)-\U'(0^-)}{2(1+r)\kd}\ - \frac{\g(1-r)}{2\sqrt{\s(1+r)}}\ \kd^{-\frac{3}{2}} = \frac{U_{-\infty}}{V}\ . \label{k_eq}
\ee
For convenience, we introduce the following notation:
\begin{align}
\st\equiv \frac{\s}{1+r},\qquad &\gt\equiv \frac{1-r}{1+r}\ \g, \\
\A\equiv \frac{\U'(0^-)-r\U'(0^+)}{1+r}\quad\text{and}\quad &\B\equiv \frac{\Us-U_{-\infty}}{V}\ .
\end{align}
We assume a wind-induced current in the water so that $\B>0$. Following \citet{young-wolfe}, we let the inverse Weber number be a small parameter, $\st\ll1$, and solve equation~(\ref{k_eq}) in the form
\be
\sqrt{\st\kd}\bigg(\B-\frac{\A}{2\kd}\bigg)=\st\kd+\frac{\gt}{2\kd},\qquad \st\ll1. 
\ee
Taking the square of this result and letting $X\equiv \st\kd$ we obtain
\be
X\bigg[\B-\frac{\st\A}{2X}\bigg]^2= \bigg[X+\frac{\st\gt}{2X}\bigg]^2,\qquad \st\ll1. \label{X_eq}
\ee
Equation (\ref{X_eq}) can be readily solved using regular perturbation theory. We seek solutions in the form
\be
X = X_0 + \st\ X_1+O\big(\st^2\big). 
\ee
We find
\be 
X_0= \B^2\qquad\text{and}\qquad X_1 = -\frac{ \A\B +\gt}{2X_0 -\B^2}\ ,
\ee
and infer 
\be
\kd^+_{\rm{neut}} = \frac{\B^2(1+r)}{\s} - \frac{\U'(0^-)-r\U'(0^+)}{(1+r)\B}- \frac{1-r}{1+r}\ \frac{\g}{\B^2} + O(\s).
\ee
For the double exponential profile introduced in \S \ref{profiles} we find
\be
\kd^+_{\rm{neut}} = \frac{1+r}{\s} - \frac{1-r(\Rone-1)/\Rtwo}{1+r} - \frac{1-r}{1+r}\ \g + O(\s),
\ee
which is a generalization of equation (5.10) in \cite{young-wolfe}.
%
%

%

\end{document}